\documentclass[]{rsos}%%%%where rsos is the template name

 \usepackage{siunitx}
\usepackage{ragged2e}
\usepackage{array}
\usepackage{cite}

\begin{document}
\begin{sloppypar}

%%%% Article title to be placed here
\title{Aircraft Ground Taxiing Deduction and Conflict Early Warning Method Based on Control Command Information}

\author{%%%% Author details
Jingchang. Zhuge$^{1}$, Huiyuan. Liang$^{2}$ and Yiming. Zhang$^{3}$, Shichao. Li$^{4}$, Xinyu. Yang$^{5}$ and Jun. Wu$^{6}$}

%%%%%%%%% Insert author address here
\address{$^{1}$Associate Professor,College of Electronic Information and Automation, Civil Aviation Univ. of China\\
$^{2}$Master’s Candidate,College of Electronic Information and Automation, Civil Aviation Univ. of China\\
$^{3}$Master’s Candidate,College of Electronic Information and Automation, Civil Aviation Univ. of China\\$^{4}$Master’s Candidate,College of Electronic Information and Automation, Civil Aviation Univ. of China\\
$^{5}$Master’s Candidate,College of Electronic Information and Automation, Civil Aviation Univ. of China\\
$^{6}$Associate Professor,College of Electronic Information and Automation, Civil Aviation Univ. of China}

%%%% Subject entries to be placed here %%%%
\subject{Engineering and technology, computer modelling and simulation}

%%%% Keyword entries to be placed here %%%%
\keywords{ Control command, Aircraft taxiing, Process deduction, Conflict early warning}

%%%% Insert corresponding author and its email address}
\corres{Jingchang Zhuge\\
\email{: 12315414@qq.com}}

%%%% Abstract text to be placed here %%%%%%%%%%%%
\begin{abstract}
Aircraft taxiing conflict is a threat to the safety of airport operations, mainly due to the human error in control command information. In order to solve the problem, The aircraft taxiing deduction and conflict early warning method based on control order information is proposed. This method does not need additional equipment and operating costs, and is completely based on historical data and control command information. When the aircraft taxiing command is given, the future route information will be deduced, and the probability of conflict with other taxiing aircraft will be calculated to achieve conflict detection and early warning of different levels. The method is validated by the aircraft taxi data from real airports. The results show that the method can effectively predict the aircraft taxiing process, and can provide early warning of possible conflicts. Due to the advantages of low cost and high accuracy, this method has the potential to be applied to airport operation decision support system.
\end{abstract}

%%%%%%%%%%%%%%%%%%%%%%%%%%%
%%%%%%%%%% Insert the texts which can accomdate on firstpage in the tag "fmtext" %%%%%

\begin{fmtext}
\end{fmtext}

%%%% Insert A head here
\maketitle

\section{Introduction}

The civil aviation industry is important to the world's economic and social development. In recent years, with the rapid increase in the number of flights, the airport operation scheduling is facing great challenges\cite{ref1, ref2}, which directly affects the safety and efficiency of airports. Air traffic controllers, as the core of operation scheduling, with the increasingly busy airport and the increasing pressure of controllers, also gradually increase the probability of controller behavior errors, which can easily lead to taxiing conflicts and other accidents and incidents in the airport. For example, On October 11, 2016, a serious runway intrusion occurred at Shanghai Hongqiao International Airport in China. Blame the tower controller for forgetting the taxiing state of the aircraft. Therefore, how to avoid accidents and incidents caused by human factors of controllers, how to reduce the pressure of controllers by verifying the control command, and how to improve or enhance the supervision and security work by using auxiliary facilities and technical methods have become problems that need to be solved urgently.

Nowadays, most hub airports in the world adopt surface surveillance systems (Primary Radar, Secondary Radar, Aerodrome Surface Multilateration Systems and ADS-B Systems) and Advanced Surface Movement Guidance and Control Systems (A-SMGCS) to solve the above problems. A-SMGCS is a comprehensive integrated system that realizes the control of aircraft in movement area through surveillance, routing planning and guidance function. In addition, Lincoln Laboratory is developing the Small Airport Surveillance Sensor (SASS), which realizes radar surveillance of small airports through the collection of aircraft images. Massachusetts Institute of Technology has developed a hazard warning system based on probability models that can warn operators of possible adverse events in the future so that measures can be taken to reduce risk\cite{ref3}. However, the surface surveillance systems and A-SMGCS are not only expensive in airport expansion, but also highly dependent on surveillance equipment. The installing and upgrading of such systems in airports require a large amount of capital investment and non-stop construction, which puts huge pressure on the operation of the airport.

The mainly terms of relevant theoretical research are as followed: the airport operation scheduling capability is improved through the flight area scheduling optimization \cite{ref4,ref5,ref6,ref7,ref8,ref9}, the aircraft taxiing route optimization \cite{ref10,ref11,ref12,ref13}, reducing emissions\cite{ref14,ref15,ref16,ref17,ref18}and the conflict detection method optimization \cite{ref19,ref20,ref21,ref22,ref23,ref24}. Zhang et al.\cite{ref25} provided a consistent method to improve the methods of determining unimpeded taxiing time, which is the reference time used for estimating taxiing delay. Brownlee et al. proposed a taxi time and uncertainty estimation system based on adaptive Mamdani fuzzy rules, and the delays due to uncertain taxi time was reduced by 10–20\% \cite{ref26}. Hasnain Ali et al. use the time-space model of statistical learning to calculate the conflict probability of the identified taxiway intersection to evaluate the conflict coefficient and the heat value of the hotspot \cite{ref27}.

All the taxiing routes in the studies mentioned above are determined by their own or others' methods, while the taxi route in this manuscript is based on control commands and is determined by air traffic controllers, which is essentially different. Although the goal of airport operation is to achieve automation and unmanned operation, the researches above ignore that in the actual control process, most airports still use manual control in airport operation scheduling for the safety of airport operations. Less attention is paid to verifying the correctness of the commands given by the controller. In addition, current aircraft taxiing conflict detection is based on the scene monitoring technology to obtain the real-time position of the aircraft\cite{ref28}. Although it has realized a certain level of early warning, the dependence on equipment and the time of early warning needs to be improved (after the controller gives the command to start taxiing, in a general sense, this is at least 10 minutes before the aircraft takes off). Therefore, there are four problems in conflict detection during taxiing: over-reliance on expensive equipment, lack of attention to the actual control process, lack of the derivation of taxiing process and the correctness verification of the control commands.

Aiming at the above problems, this paper proposes an aircraft taxiing deduction and conflict early warning method based on control command information. By analyzing the ADS-B historical data of the aircraft in real operation, the taxiing speed interval of the aircraft at different positions on the taxiway is obtained. Combined with the control command information issued by the controller, the taxiing path of the aircraft can be deduced in advance. According to the designed aircraft taxiing conflict detection and early warning method, the probability of conflict between aircraft can be obtained, and different types and different levels of conflict early warning can be realized. The method proposed in this paper can realize the deduction of the ground taxiing process of the aircraft and the early warning of the aircraft taxiing conflict without the need for airport monitoring equipment and shutdown construction. In addition, air traffic controllers can reschedule the taxiing time of the next taxiing aircraft according to the calculated taxiing path, and arrange idle taxiing routes, which is beneficial to reduce aircraft waiting time and reduce fuel consumption. In general, the research in this paper has a positive effect on avoiding aircraft taxiing conflicts, reducing the pressure on controllers, reducing the cost of airport construction and renovation, and improving the efficiency of airport operations.

\section{Description of the Aircraft Taxiing Conflict Problem}
Aircraft must go through the ground taxi process before takeoff and after landing. This process is completed in the airport flight area, and the main connection channel is the taxiway network. The taxiing phase is often the most complex phase of an aircraft's operation at an airport. The taxiing of the aircraft is accomplished through the radiotelephony communication between the controller and the pilot. After the pilot receives the controller's control command information, he will taxi strictly according to the route in the control command information through the ground signs. There are three potential aircraft taxiing conflicts during the taxiing process, as shown in figure 1.

\begin{figure}[ht]
%\centering\includegraphics[width=2.5in]{xxxxxx.eps}
%%% where xxxxxx name represents "figurename.eps"
\centering\includegraphics[width=5in]{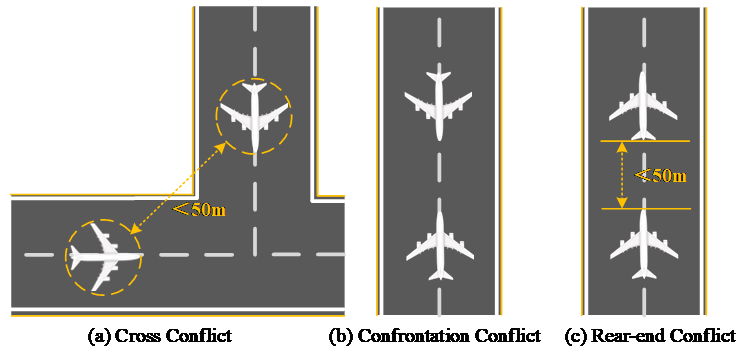}
\caption{Schematic diagram of typical taxi conflict}
\label{fig1}
\end{figure}
1.	Cross conflict

When two aircraft cross their original routes and have met at the same taxiway intersection and the safety separation is not met at this time, a cross conflict may occur. 

2.	Confrontation conflict

When two aircraft taxi opposite each other on the same section and there is no other intersection, a confrontation conflict will occur. 

3.	Rear-end conflict

When two aircraft are taxiing in the same direction on the same section, a rear-end conflict may occur when the safety separation is not met at this time.
The taxiing process of the aircraft must meet certain rules of operation, which mainly include:

1.	The taxiing route is uniquely determined and in a single direction. It must be taxied along the path specified by the command, but the flight crew can adjust the taxiing state (start, accelerate, constant speed, decelerate, stop, turn) according to the taxi-way parameters (turning position, intersection position, straight length, etc.).

2.	All aircraft should maintain at least a specified safe separation. When two or more aircraft are taxiing, the aircraft behind are not allowed to exceed the aircraft in front, and the minimum separation is not allowed to be less than \SI{50}{m}.

3.	The aircraft should taxi within the designated taxiing speed range, which is not allowed to exceed \SI{50}{km/h}.

Therefore, the problem to be solved in this paper can be described as: how to provide a method for aircraft ground taxiing deduction and conflict warning based on control command information, simulate the taxiing process of the aircraft in advance and detect the aircraft taxiing conflict and give early warning before the controller issues the control command information.

\section{Deduction Method of Aircraft Taxiing Process}
\subsection{Data Preprocessing}
All operating aircraft at the target airport are equipped with the ADS-B system. The ADS-B system is an air traffic control surveillance system based on global satellite positioning technology. It conducts traffic surveillance and information transmission based on air-air and air-ground data links. According to the different frequencies and codes transmitted by the transponders loaded on the aircraft, there are mainly A Mode, C Mode and S Mode. At present, most countries and organizations in the world, such as China, the United States and the European Union, stipulate that the airports and aircraft currently operating must be equipped with the ADS-B system and support the S Mode.

Receive ADS-B messages such as$ Data Cat=01$, $Data Cat=05$ or $07$,$ Data Cat = 09$, $Data Cat = 38$ of historical data in ADS-B system of 13879 aircrafts within one month in target airport through SBS-3 receiver. Based on the STANDARD OF DO-260A, the ME field in the message is analyzed and the TYPE value is extracted, and the 24-bit ICAO address code, taxiing speed, heading angle, and geographic location coordinates of all operating aircraft in the target airport can be obtained.

Among them, the geographic location information in the ADS-B system generates geographic coordinates under the World Geodetic System—1984 Coordinate System (WGS-84 coordinate system), including geodetic latitude $B$, geodetic longitude $\lambda$, and altitude $H$. The WGS-84 coordinate system is a geocentric coordinate system whose origin is the earth's center of mass. The Z-axis of the coordinate system points to the Conventional Terrestrial Pole (CTP) defined by the BIH (International Time Service Agency), and the X-axis points to the intersection of the zero-degree meridian plane of BIH 1984.0 and the CTP equator. The commonly used earth rectangular coordinate system is Earth-Centered Earth-Fixed (ECEF). ECEF is a coordinate system that rotates with the rotation of the earth. Convert the coordinates ($B$, $\lambda$, $H$) in the WGS-84 coordinate system to the coordinates ($x$, $y$, $z$) in the ECEF coordinate system. The transformed ECEF coordinates ($x$, $y$, $z$) are as follows.
\begin{align}\label{3.1}
\begin{split}
\left\{\begin{matrix} 
  x=(C + H)\text{cos}B {cos}\lambda\\  
  y=(C + H)\text{cos}B{sin}\lambda \\  
  z=[C(1-e^{2}+H )]\text{sin}B 
\end{matrix}\right. 
\end{split}
\end{align}

Among them,

\begin{align}\label{3.2}
\begin{split}
C=\frac{E_{q}}{\left(1-e^{2} \sin ^{2} B\right)^{\frac{1}{2}}} 
\end{split}
\end{align}

In formula (3.2), $E_q$ is the equatorial radius, and e is the eccentricity of the earth.

However, there are the following problems in parsing the ADS-B data of the target airport: (1) The ADS-B system has the phenomenon of co-channel interference, so some ADS-B data is missing or inconsistent; (2) The parsed data is high frequency and large The data does not directly give the required taxiing information of the aircraft, and there are data irrelevant to the re-search purpose; (3) The source of the aircraft geographic location data is the global navigation satellite system, which leads to certain deviations in the aircraft data. In summary, the original data needs to be analyzed and processed.

First of all, due to the deviation of the geographic location information in the acquired data of the aircraft, there is a phenomenon that the geographic coordinates are located on the taxiway but the taxiing speed value is far greater than \SI{50}{km/h}. The value of the maximum taxiing speed of \SI{50}{km/h} on the airport surface stipulated by the Civil Aviation Administration of China (CAAC) is excluded. Since the aircraft taxiing speed interval established in this paper takes into account the aircraft taxiing waiting constraint, it is not necessary to remove the extremely low taxiing speed value.

Second, every aircraft equipped with the ADS-B system has a 24-bit ICAO address code. It is the code assigned to the S Mode transponder and the unique identification code of the aircraft registered with the International Civil Aviation Organization (ICAO). Through the 24-bit ICAO address code, the aircraft type and registration number can be queried and matched. Based on this, an information table of all operating aircrafts belonging to the target airport (airline affiliated, 24-bit ICAO address code, aircraft type and aircraft registration number) is established.

Finally, the taxiing data of the aircraft are classified, extracted and summarized according to the taxiway name. According to the frequency of use of taxiways in the target airport, eight high-frequency taxiways are selected. By defining the geographic coordinate range of each taxiway, the geographic coordinate interval of each taxiway is defined, and then the taxiing speed of each aircraft passing through the same taxiway (the geographic coordinate passing through is within the geographic coordinate interval of each taxiway) is extracted and summarized. Through the Kolmogorov-Smirnov test, it is judged that the taxiing speed data conforms to the Gaussian distribution. In order to more accurately express the range of the reliability of the taxiing speed, and for the subsequent experimental verification by using the comparison method, an interval is determined for the set of taxiing speed values of each taxiway according to the Pauta criterion, and this type of interval is named as Pauta taxiing speed interval ( $\Delta$$Vk-Pauta$, $k$ is the taxiway number).

\subsection{Constraint relationship of aircraft taxiing speed}

The speed of the aircraft when taxiing in the airport is not an arbitrary value, but has a natural and reasonable range. The reason is that the aircraft will be affected by the airport operating rules when taxiing. Therefore, the constraint relationship of the taxiing speed of the aircraft is analyzed based on the real historical taxiing speed data of the aircraft in the target airport.

Take an airport in North China as an example to analyze the constraint relation of the taxiing speed of the aircraft. 13,879 inbound and outbound flights were counted in one month in August 2019. According to the analysis of historical flight information at the airport, combined with China's latest RECAT-CN aircraft classification standard, it can be learned that there are no J-type and L-type aircraft in the airport flights, while B-type, C-type and M-type aircraft are operated in domestic and international flights. The proportion of operating aircraft types is shown in figure 2.
\begin{figure}[ht]
%\centering\includegraphics[width=2.5in]{xxxxxx.eps}
%%% where xxxxxx name represents "figurename.eps"
\centering\includegraphics[width=5in]{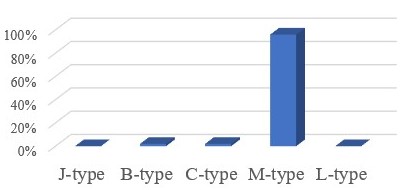}
\caption{Scale diagram of operating aircraft type}
\label{fig2}
\end{figure}

According to the proportion of aircraft types in the target airport, ignoring the J-type, B-type, C-type and L-type aircraft, six common M-type aircraft in the target airport are selected: A319, A320, A321, B737-700, B737-800 and E190 as experimental targets. The eight high-frequency taxiways selected in the previous section are used as the experimental taxiways, and the selections are as follows. Since taxiway B is a long-distance parallel taxiway, and due to the large flow, the taxiing speed value of the aircraft passing through each section of the taxiway fluctuates greatly, so the taxiway B is divided into five sections according to the terrain structure. Defined as B-1 (between B1 and B2), B-2 (between B2 and B3), B-4 (between B3 and B4), B-6 (between B4 and B4) Between B6), B-8 (between B6 and B8), and then select the remaining three taxiways (C, N, and P) with higher traffic. A total of eight taxiways are selected.
Finally, through the data processing steps of parsing ADS-B messages, coordinate transformation, removing outliers, establishing aircraft information table and establishing Pauta taxiing speed interval in Section 3.1, 136222 pieces of aircraft taxiing data are finally obtained.

\subsubsection{The Constraint Relationship between the Length of the Taxiway and the Taxiing speed of the Aircraft}
Firstly, analyze the constraint relationship between the length of the taxiway and the taxiing speed of the aircraft when the aircraft is taxiing. When the same type of aircraft taxis on different taxiways, the specific data of the length of the taxiway and the average taxiing speed during taxiing are shown in table 1.

\begin{table}[ht]\centering
\caption{The data of taxiway length and taxiing speeds for aircraft of the same type during taxiing}%%%Table caption goes here
\label{table1}
\centering
   \begin{tabular}{m{3.5cm}<{\centering}m{0.8cm}<{\centering}m{0.8cm}<{\centering}m{0.8cm}<{\centering}m{0.8cm}<{\centering}m{0.8cm}<{\centering}m{0.8cm}<{\centering}m{0.8cm}<{\centering}m{0.8cm}<{\centering}}
    \hline
    Taxiway &B-1 &B-2 &B-4 &B-6 &B-8 &C &N &P \\
    \hline
    Taxiway length ($m$) &200 &545.5 &568 &1196 &1063.5 &1196 &561.75 &401.75 \\
    Average Taxiing Speed ($kn$) &10.10 &11.20 &15.67 &12.74 &11.37 &12.06 &14.71 &15.13 \\\hline
\end{tabular}
\end{table}%%%End of the table

The scatterplot of taxiway length and taxiing speed of the aircraft is shown in figure3.

\begin{figure}[ht]
%\centering\includegraphics[width=2.5in]{xxxxxx.eps}
%%% where xxxxxx name represents "figurename.eps"
\centering\includegraphics[width=5in]{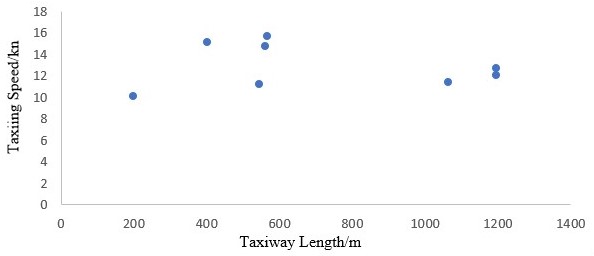}
\caption{Scatter plot of taxiway length and taxiing speed.}
\label{fig3}
\end{figure}

Through the Kolmogorov-Smirnov test, it is judged that the taxiway length and taxiing speed data follow the Gaussian distribution. Coupled with the judgment of the scatter diagram in figure 3. The relationship between taxiway length and average taxiing speed can be verified by constructing the Pearson correlation coefficient.

Pearson correlation coefficient formula is as follows.
\begin{align}\label{3.3}
\begin{split}
r=\frac{\sum_{i=1}^{n}\left(X_{i}-\bar{X}\right)\left(Y_{i}-\bar{Y}\right)}{\sqrt{\sum_{i=1}^{n}\left(X_{i}-\bar{X}\right)^{2}} \sqrt{\sum_{i=1}^{n}\left(Y_{i}-\bar{Y}\right)^{2}}}
\end{split}
\end{align}

Calculated by formula (3.3), the Pearson correlation coefficient between taxiway length and average taxiing speed is $-0.12$. According to the definition of the Pearson correlation coefficient, the absolute value of the Pearson correlation coefficient of the two variables is in the interval of $0-0.2$, which is a very weak correlation or no correlation. And through test, it is found that the length of the taxiway has no significant effect on the taxiing speed, and the conclusion is consistent with the conclusion drawn by the Pearson correlation coefficient. It can be proved that there is no correlation between the length of the taxiway on which the aircraft is taxiing and the taxiing speed of the aircraft. Therefore, in the subsequent deduction of the taxiing process of the aircraft and the detection and early warning of the taxiing conflict of the aircraft, the influence on the taxiing speed of the aircraft under this constraint can be ignored.

\subsubsection{The Constraint Relationship Between Aircraft Type and Taxiing speed of the Aircraft}

Secondly, analyze the constraint relationship between the aircraft type and the taxiing speed of the aircraft when the aircraft is taxiing. In the same way, the three elements representing the six aircraft types (the maximum take-off weight of the air-craft, the unladen weight of the aircraft and the maximum thrust of the aircraft) and their Pearson correlation coefficients will be used to verify the relationship between the two.

Taking B-8 taxiway as an example, the maximum takeoff weight, empty weight, maximum thrust and average taxiing speed of six aircraft types are obtained by querying relevant aircraft information and classifying and summarizing aircraft taxiing data after data processing. Then, Pearson correlation coefficient was constructed by combining the three elements representing aircraft type and the taxiing speed. As shown in table 2.

\begin{table}[ht]\centering
\caption{Data representing three elements of aircraft type and taxiing speeds in the B-8 taxiway}%%%Table caption goes here
\label{table2}

\centering
    \begin{tabular}{m{4cm}<{\centering}m{2cm}<{\centering}m{2cm}<{\centering}m{2cm}<{\centering}m{2cm}<{\centering}}%%%The number of columns has to be defined here
    \hline
    B-1 taxiway &Aircraft maximum take-off weight (kg) &Aircraft unloaded weight (kg) &Aircraft maximum thrust (kN) &Aircraft average taxiing speed (kn) \\
    \hline
    A320/A20N &78000 &42400 &292.6 &8.88 \\
    B737 &79010 &37648 &232 &11.5 \\
    B738/B739 &78245 &41413 &232 &9.88 \\
    A321/A21N &93500 &48200 &292.6 &11.46 \\
    A319 &75500 &40600 &267.2 &14.69 \\
    E190/E195 &51000 &28080 &164.6 &9.56 \\
    Pearson correlation coefficients constructed respectively with the aircraft taxiing speed &0.26 &0.22 &0.32 & \\\hline
\end{tabular}

\end{table}%%%End of the table

In the same way, the Pearson correlation coefficients of the three elements representing the aircraft type on the eight taxi-ways and the taxiing speed of the aircraft are shown in table 3.

\begin{table}[ht]\centering
\caption{Data representing three elements of aircraft type and taxiing speeds in the B-8 taxiway}%%%Table caption goes here
\label{table3}
    \begin{tabular}{m{3.5cm}<{\centering}m{3cm}<{\centering}m{3cm}<{\centering}m{3cm}<{\centering}}%%%The number of columns has to be defined here
    \hline
    Taxiway &Pearson correlation coefficient of the aircraft maximum take-off weight and aircraft taxiing speed &Pearson correlation coefficient of the aircraft unloaded weight and aircraft taxiing speed &Pearson correlation coefficient of the aircraft maximum thrust and the aircraft taxiing speed \\
    \hline
    B-1 &0.00 &0.03 &0.05 \\
    B-2 &0.42 &0.40 &0.35 \\
    B-4 &0.39 &0.42 &0.44 \\
    B-6 &-0.37 &-0.39 &-0.42 \\
    B-8 &0.26 &0.22 &0.32 \\
    C &0.43 &0.45 &0.47 \\
    N &0.37 &0.39 &0.38 \\
    P &0.00 &0.03 &0.05 \\
    Pearson correlation coefficients(average) &0.28 &0.29 &0.31 \\
   Pearson correlation coefficients(median) &0.37 &0.39 &0.36 \\
  \hline
\end{tabular}
\end{table}%%%End of the table

According to the correlation definition of Pearson correlation coefficient, the absolute value of Pearson correlation coefficient of two variables is in the interval of $0.2-0.4$, which belongs to weak correlation. It can be seen from table 3 that the Pear-son correlation coefficients (average and median) of the three elements representing the aircraft type and the structure of the air-craft taxiing speed are all within the interval of $0.2-0.4$. And after t test, it is found that the three elements of the representative model have no significant effect on the taxiing speed of the aircraft, and the conclusion is consistent with the conclusion drawn by the Pearson correlation coefficient. Therefore, it is considered that under the same model, there is a weak correlation between the aircraft type and the taxiing speed of the aircraft in the same taxiway, and there is no strong correlation. Therefore, in the subsequent deduction of the taxiing process of the aircraft and the detection and early warning of the taxiing conflict of the aircraft, the influence on the taxiing speed of the aircraft under this constraint can be ignored.

\subsubsection{The Constraint Relationship between the Taxiing Date (Weekdays and Weekends) and the Taxiing Speed of the Aircraft}

Furthermore, analyze the constraint relationship between the date (weekdays and weekends) of the aircraft taxiing and the taxiing speed of the aircraft. By classifying and summarizing the aircraft taxiing data after data processing, the average taxiing speed values on weekdays and weekends in the eight taxiways can be obtained as shown in table 4.

\begin{table}[ht]
\caption{The data of taxiway length and taxiing speeds for aircraft of the same type during taxiing}%%%Table caption goes here
\label{table4}
 \begin{tabular}{m{5cm}<{\centering}m{4cm}<{\centering}m{4cm}<{\centering}}%%%The number of columns has to be defined here
\hline
Taxiway &Average taxiing speed during weekdays (\text{kn}) &Average taxiing speed during weekends (\text{kn}) \\
\hline
    B-1 &7.87 &9.77 \\
    B-2 &8.49 &10.70\\
    B-4 &15.66 &15.44 \\
    B-6 &12.58 &12.01 \\
    B-8 &9.44 &8.34 \\
    C &6.01 &6.13\\
    N &13.78&13.36 \\
    P &14.96&14.45 \\
   Pearson correlation coefficient of average aircraft taxiing speed on weekdays and week-ends &0.94 \\
    \hline
\end{tabular}
\end{table}%%%End of the table

From table 4, it can be clearly seen that the taxiing speed value of the aircraft on each taxiway during the weekend is greater than the taxiing speed value on the weekday, indicating that the taxiing speed of the aircraft on the weekend is faster than the taxiing speed of the aircraft on the weekday. And from the definition of Pearson correlation coefficient, it can be known that the absolute value of Pearson correlation coefficient of two variables is in the interval of $0.8-1$, which is a very strong correlation. In table 4, the Pearson correlation coefficient of the average taxiing speed on weekdays and weekends is 0.94, which is between 0.8 and 1. And after the t test, the relationship between the two variables was found to be very significant. Therefore, it is considered that the average taxiing speed of the aircraft on weekdays and weekends has a strong correlation. Therefore, in the subsequent deduction of the aircraft taxiing process and the detection and early warning of aircraft taxiing conflicts, the impact of weekdays and weekends on the aircraft taxiing speed should be fully considered.

\subsubsection{The Constraint Relationship between the Time (Day and Night) and the Taxiing Speed of the Aircraft}

Finally, analyze the constraint relationship between the time (day and night) of the aircraft taxiing and the taxiing speed of the aircraft. By classifying and summarizing the aircraft taxiing data after data processing, one day is divided into seven time periods, and the average taxi speed values of the seven time periods in the eight taxiways are shown in table 5.

\begin{table}[ht]
\caption{Average taxi speeds for seven time periods in eight taxiways}%%%Table caption goes here
\label{table5}
\centering
 \begin{tabular}{m{1cm}<{\centering}m{1.4cm}<{\centering}m{1.4cm}<{\centering}m{1.4cm}<{\centering}m{1.4cm}<{\centering}m{1.4cm}<{\centering}m{1.4cm}<{\centering}m{1.4cm}<{\centering}}%%%The number of columns has to be defined here
\hline
Taxiway &00:00:00-5:59:59 (kn)&6:00:00-8:59:59 (kn) &9:00:00-11:59:59 (kn)&12:00:00-14:59:59 (kn)&15:00:00-17:59:59 (kn)&18:00:00-20:59:59 (kn)&21:00:00-23:59:59 (kn)\\
\hline
    B-1 &6.6613 &7.7595&8.4939&11.2083&7.5886&6.9462&7.2667\\
    B-2 &9.4956&11.3164&10.1598&11.5290&10.3655&10.1732&7.0609\\
    B-4 &14.0335 &14.7131&15.0969&16.2706&15.0286&14.3831&14.5460 \\
    B-6 &11.8357 &11.1401&12.1447&12.8346&12.0711&11.7268&11.6429 \\
    B-8 &8.9176 &9.4636&9.5489&13.2683&10.9994&9.3143& 9.2469\\
    C &9.9141&10.3548&10.9352&10.8746&11.0982&10.4523&10.3079\\
    N &12.0792&12.8435&13.9674&14.7938&12.7076&12.7891& 12.8783\\
    P &12.8841&13.8592&15.2678&14.8428&12.9716&12.4286& 12.1253\\
    \hline
\end{tabular}
\end{table}%%%End of the table

From table 5, it can be clearly seen that the average taxiing speed of the three time periods (00:00:00-5:59:59, 18:00:00-20:59:59and 21:00:00-23:59:59) are both at night. It is significantly lower than that of during the day. This complies with Article 299 of the "China Civil Aviation Air Traffic Management Regulations" CCAR-93TM-R5: aircraft taxiing, air taxiing and towing shall comply with the following regulations: (6) When taxiing or towing at night, the navigation should be turned on. Lights and taxi lights, or intermittent use of landing lights, the aircraft must taxi at low speed at night. Therefore, in the subsequent deduction of the taxiing process of the aircraft and the detection and early warning of the taxiing conflict of the aircraft, the in-fluence of the day and night on the taxiing speed of the aircraft should also be fully considered.

\subsection{ Aircraft taxiing assumptions}

In order to facilitate the deduction of the essential attributes and internal connections of key technologies such as aircraft taxiing and subsequent conflict early warning. It can not only objectively reflect the essence of the taxiing process of the aircraft, but also facilitate discussion and processing. Three aircraft taxiing assumptions are presented here.

\subsubsection{ Information Completeness Assumption}

The complete, clear and accurate taxiing control command information issued by the controller is unique, that is, the unique routing information of the aircraft taxiing is known from the moment the controller completes the command. It is considered that the basic information, including the flight plan of the aircraft and the taxiing path of the aircraft is complete, determined and known before the aircraft taxis.

\subsubsection{Aircraft taxiing waiting Assumptions}

With the occurrence of traffic congestion and delays in the airport flight area, the aircraft may be evasive and waiting when taxiing. However, the taxiing speed interval defined in this paper already contains a large number of taxiing speed values of aircraft waiting time after data processing in Section 3.1. Therefore, it can be considered that the taxiing of the aircraft defined in this paper is the taxiing after having been constrained by waiting.

\subsubsection{Special Vehicle Operation Assumptions}

The focus of this paper is on aircraft taxiing deduction and conflict early warning research. Therefore, when the aircraft is taxiing, some operational constraints are simplified, and the impact of airport special vehicles on aircraft taxiing is not considered.

\subsection{The Method of Taxiing Process Deduction of the Aircraft}

At present, most of the researches on predicting the taxiing process of the aircraft are separated from the simulation of the taxiing of the aircraft in the real control environment. At the same time, there is no way to deduce the taxiing process of the aircraft based on the control command information after the controller issues the control command. Therefore, in order to predict the taxiing process of the aircraft more reasonably, this section proposes a deduction method for the taxiing process of the aircraft based on the control command. The purpose is to provide a solution for simulating the aircraft taxi routing plan in advance, after the controller has given control command, before the aircraft begins to actually taxi out/in. A new paradigm of aircraft taxiing process deduction method is proposed.

The current aircraft trajectory prediction technology in airports is uncertain and random. The accuracy of predictions is affected by various sources of error, including the effect of wind, the effect of modeling errors, and the effect of positional errors. Furthermore, even when the effects of various errors are minimized, trajectory prediction cannot account for the controls taken by pilots and controllers. In view of this situation, this paper proposes an aircraft taxiing process deduction method based on control command to deduce the speed and time of the aircraft taxiing in advance when the aircraft arrives at each taxiway in the future. Since the deduction method proposed in this section strictly follows the taxiing path in the control command, the prediction of the aircraft trajectory in the airport is deterministic and not random.

In the existing research on aircraft taxiing prediction, it can be found that most of them define the taxiing speed of the air-craft as a quantitative or uniform variable, but in the actual taxiing process of the aircraft is not a simple uniform speed process or uniform acceleration and deceleration process, but a process of random change. In order to more reasonably simulate and predict the speed of the aircraft taxiing on different taxiways in the flight area, this section defines the aircraft taxiing speed as an interval variable, which is closer to the real taxiing process of the aircraft. Through various constraints that affect the taxiing speed of the aircraft in the previous section, different speed ranges of different aircrafts when taxiing on different taxiways can be obtained. In order to define a more accurate taxiing speed interval, the following is a re-classification, sorting and summary of the taxiing speed interval of the aircraft.

From the previous section, it is known that the correlation between the two constraints of aircraft type and taxiway length and the taxiing speed of the aircraft is weak or irrelevant, so it is ignored when establishing the taxiing speed interval. The date of taxiing (weekdays and weekends) and the time of taxiing (day and night) are strongly related to the taxiing speed of the aircraft, so they are reserved. To sum up, the factors affecting the taxiing of the aircraft are: the date of taxiing (weekdays and weekends) and the time of taxiing (day and night). Therefore, it is necessary to classify, extract and summarize the original data of the taxiing speed of the aircraft. First, the aircraft taxiing speed data on each taxiway is classified and extracted into four parts (weekday day, weekday night, weekend day and weekend night) according to the taxiing time of the aircraft. Second, repeat the data processing of the five steps in Section 3.1. Finally, four taxiing speed intervals are obtained on each taxiway, which are collectively referred to as time-type taxiing speed intervals ($\Delta$$Vk-time$), which are defined as $\Delta$$Vk_{wdd}$, $\Delta$$Vk_{wdn}$, $\Delta$$Vk_{wed} $and $\Delta$$Vk_{wen}$, respectively. The specific flow of the deduction method of the aircraft taxiing process is shown in figure 4.

\begin{figure}[ht]
%\centering\includegraphics[width=2.5in]{xxxxxx.eps}
%%% where xxxxxx name represents "figurename.eps"
\centering\includegraphics[width=5in]{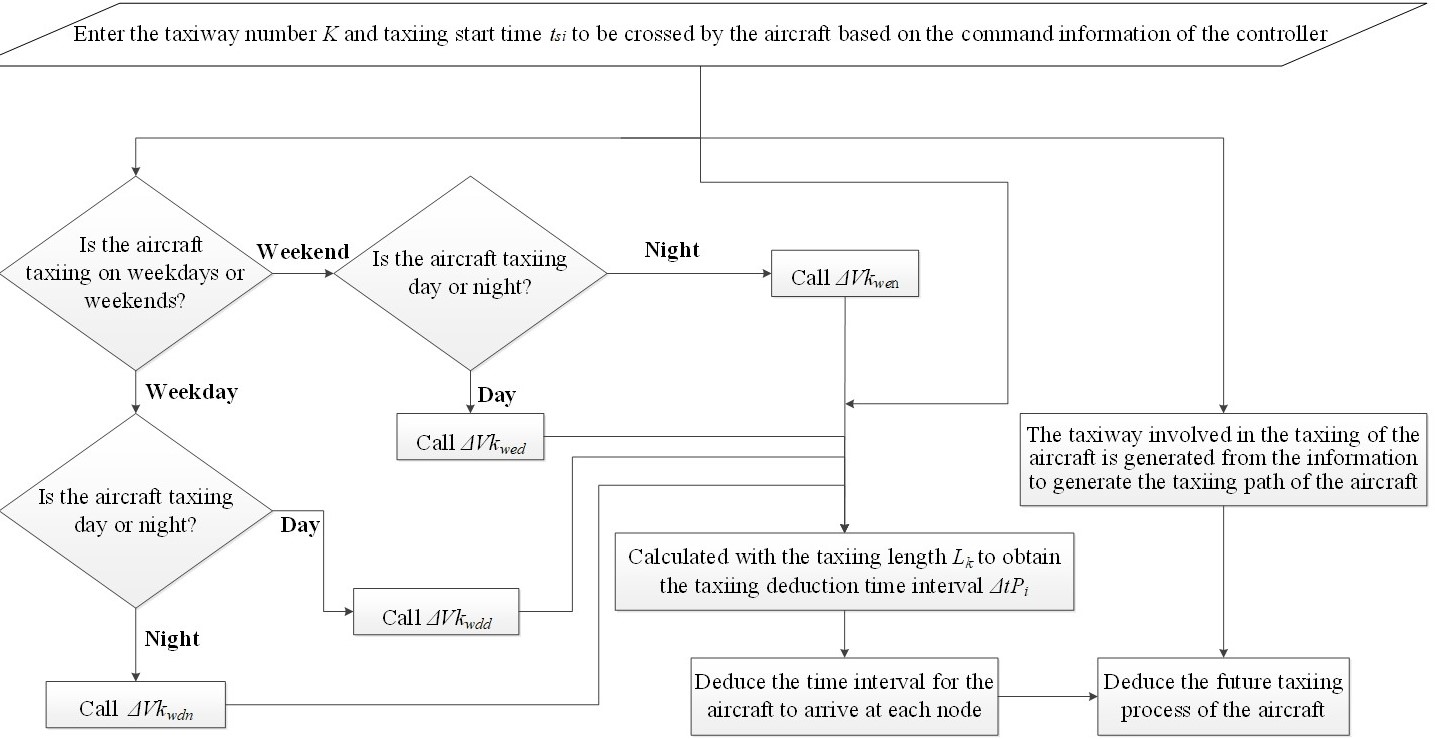}\caption{The method flow chart of deduce taxiing process of the aircraft}
\label{fig4}
\end{figure}

The specific steps of the deduction method of the aircraft taxiing process are as follows.

Step 1: Enter the taxiway $k$ in the control command and the taxiing time $t_{si}$ of the aircraft, where $k$ represents different taxi-way numbers. $t_{si}$ represents the time when the aircraft $P_i$ starts taxiing at the initial taxiing point. Among them, for the departing aircraft, the start taxiing time is equivalent to the time it takes to start taxiing from the taxiway at the parking position. For in-coming aircraft, the start taxi time is equivalent to the time the aircraft leaves the runway and enters the rapid exit taxiway.

Step 2: Determine whether the taxiing time $t_{si}$ of the aircraft is on a weekday or a weekend and whether it is during the day or night. If it is during the daytime on a weekday, $\Delta$$Vk_{wdd}$. If it is on a weekday night, call $\Delta$$Vk_{wdn}$. If it is during the daytime on a weekend, call  $\Delta$$Vk_{wed}$. If it is a weekend night, call  $\Delta$$Vk_{wen}$.

Step 3: Calculate the taxiing deduction time interval $\Delta$$tp_i$ when the aircraft Pi arrives at the taxiway $k$ using the called $\Delta$$Vk-time$ and the taxiway length $L_k$ and the time $t_{si}$ when the aircraft starts to taxi.

\begin{align}\label{3.4}
\begin{split}
\Delta t P_{i}=\frac{L_{k}}{\Delta V k-t i m e}+t_{s i}
\end{split}
\end{align}

In formula (3.4),$L_k$ represents the length of each taxiway, and $\Delta$$Vk-time$ is the time-type taxiing speed interval.

\subsection{ Simulation Experiment Verification of the Deduction Method of Aircraft Taxiing Process }

In order to verify the validity and accuracy of the proposed method for the flight taxiing process, a simulation experiment is carried out on a target airport in North China. If the verification is passed, the research results can be applied to other airports. Of course, it is impossible to generalize between airports, which needs to be analyzed according to the specific operation data of the airport.

\subsubsection{ Simulation Experiment Verification Object}

The target airport operates 24 hours a day and has two runways \SI{34}{L} and \SI{34}{R} from south to north, with lengths of 3600 meters and 3200 meters respectively. Currently, both runways can be used for take-off and landing. Due to the large number of intersections on the \SI{34}{L} runway and the connected taxiway areas, and the apron is a combination of self-taxiing and single-position taxiways, the situation of possible conflicts is more complicated and more representative. Therefore, the \SI{34}{L} runway and the connected taxiway area are used as the simulation experiment area for deducing the taxiing process of the aircraft and for the subsequent detection and early warning of the aircraft taxiing conflict. The real scene of the selected area is shown in figure 5.

\begin{figure}[ht]
%\centering\includegraphics[width=2.5in]{xxxxxx.eps}
%%% where xxxxxx name represents "figurename.eps"
\centering\includegraphics[width=5in]{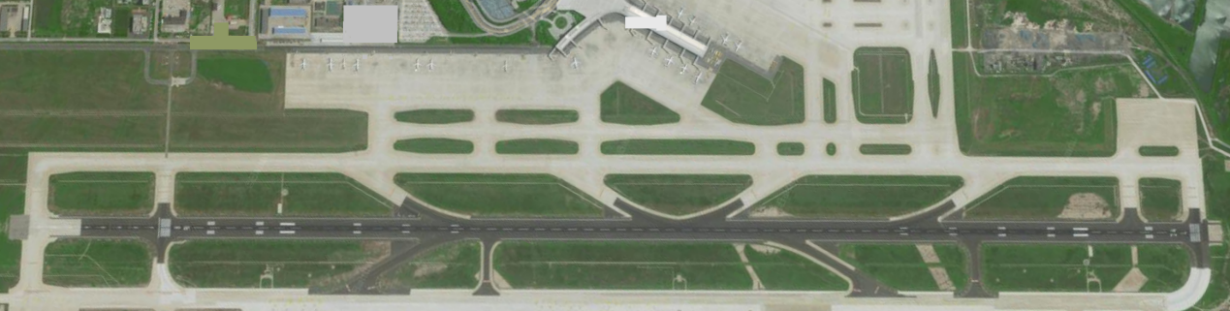}
\caption{The realistic picture of selected area}
\label{fig5}
\end{figure}

\subsubsection{Simulation experiment to verify and deduce the taxiing process of the aircraft}

In order to verify the rationality and effectiveness of the method for deriving the taxiing process of the aircraft, it is necessary to conduct verification and analysis through simulation experiments. By comparing the time interval between the deduced aircraft arriving at each node position of the taxiway and the real aircraft arriving at each taxiway node position, the fitting degree of the two is observed, and the deduction method of the aircraft taxiing process is verified by simulation experiments.

Select a real control routing command in the target airport, denoted as Command A. By querying the established information table of all operating aircraft belonging to the target airport, it can be obtained that the model is A320, and the aircraft registration number is B-300T. The taxiways involved in the aircraft taxiing in the control order are P, C, C6 and B-8 in sequence. The controller issued the command during the daytime on the weekend, so each section of the taxiway that the aircraft passed through called its own $\Delta$$Vk_{wed}$ as the taxi speed interval for the deduction. The specific taxiing path of the aircraft is shown in figure 6.

\begin{figure}[ht]
%\centering\includegraphics[width=2.5in]{xxxxxx.eps}
%%% where xxxxxx name represents "figurename.eps"
\centering\includegraphics[width=5in]{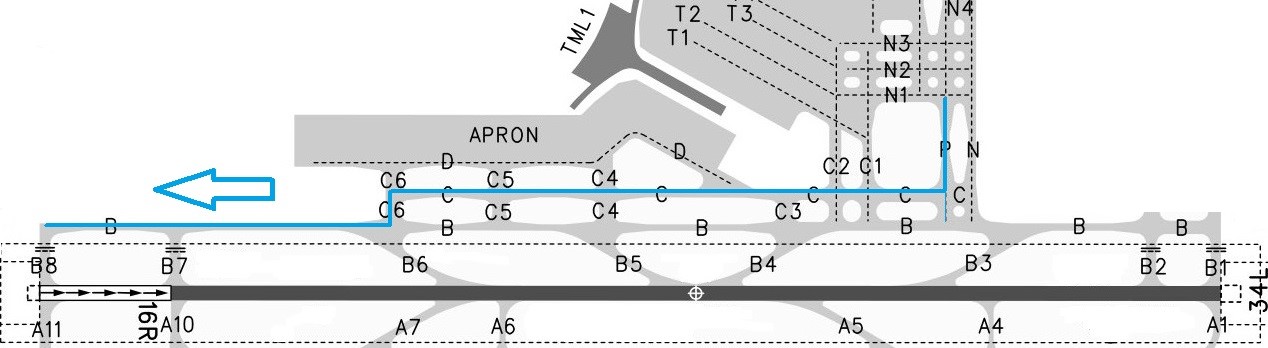}
\caption{The specific taxiing path of aircraft in control command A}
\label{fig6}
\end{figure}

The taxiing process of the aircraft in the Command A is deduced by the deduction method of the aircraft taxiing process proposed in Section 3.4, and the specific deduction taxiing data is shown in table 6.

\begin{table}[ht]
\caption{ Deduced taxiing data of aircraft in control command A}%%%Table caption goes here
\label{table6}
\centering
 \begin{tabular}{m{1cm}<{\centering}m{1.4cm}<{\centering}m{1.4cm}<{\centering}m{1.4cm}<{\centering}m{1.4cm}<{\centering}m{1.4cm}<{\centering}m{1.4cm}<{\centering}m{1.4cm}<{\centering}}%%%The number of columns has to be defined here
\hline
Number&Corresponding taxiway number&Real time$/s$ &$tp_{min}/s$ &$tp_{max}/s$& $\Delta$ $tp_{ij}/s$ &Midpoint of Interval$/s$ &Deviation \\
\hline
    B-1 &6.6613 &7.7595&8.4939&11.2083&7.5886&6.9462&7.2667\\
    B-2 &9.4956&11.3164&10.1598&11.5290&10.3655&10.1732&7.0609\\
    B-4 &14.0335 &14.7131&15.0969&16.2706&15.0286&14.3831&14.5460 \\
    B-6 &11.8357 &11.1401&12.1447&12.8346&12.0711&11.7268&11.6429 \\
    B-8 &8.9176 &9.4636&9.5489&13.2683&10.9994&9.3143& 9.2469\\
    C &9.9141&10.3548&10.9352&10.8746&11.0982&10.4523&10.3079\\
    N &12.0792&12.8435&13.9674&14.7938&12.7076&12.7891& 12.8783\\
    P &12.8841&13.8592&15.2678&14.8428&12.9716&12.4286& 12.1253\\
    \hline
\end{tabular}
\end{table}%%%End of the table

The deduction effect of the aircraft taxiing process in Command A is shown in figure 7.

\begin{figure}[ht]
%\centering\includegraphics[width=2.5in]{xxxxxx.eps}
%%% where xxxxxx name represents "figurename.eps"
\centering\includegraphics[width=4.5in]{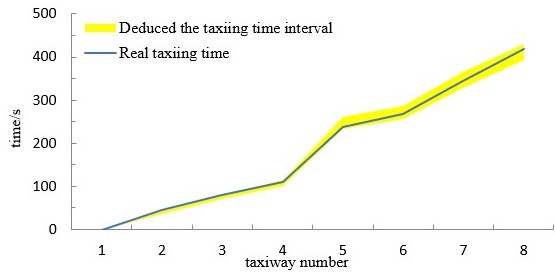}
\caption{The effect diagram of the deductive taxiing process of aircraft in Command A}
\label{fig7}
\end{figure}

In figure 6, the abscissa represents the nodes passed by the aircraft on the taxiway, and the ordinate represents the accumulated taxiing time of the aircraft. The yellow area represents the deduced taxiing time interval of the aircraft, and the solid line in the yellow area represents the actual taxiing time of the aircraft.

In the same way, another real control routing command in the target airport is selected and recorded as Command B. By querying the established information table of all operating aircraft belonging to the target airport, it can be obtained that the model is A321, and the aircraft registration number is B-6906. The taxiways involved in the aircraft taxiing in the control command are P, B-4, B-6 and B-8 in sequence. The controller gave the command at night on a weekday, so each segment of the taxiway that the aircraft passes through calls its own $\Delta$$Vk_{wdn}$ as the taxiing speed interval. figure 8 shows the specific taxiing path of the aircraft in Command B.

\begin{figure}[ht]
%\centering\includegraphics[width=2.5in]{xxxxxx.eps}
%%% where xxxxxx name represents "figurename.eps"
\centering\includegraphics[width=4.5in]{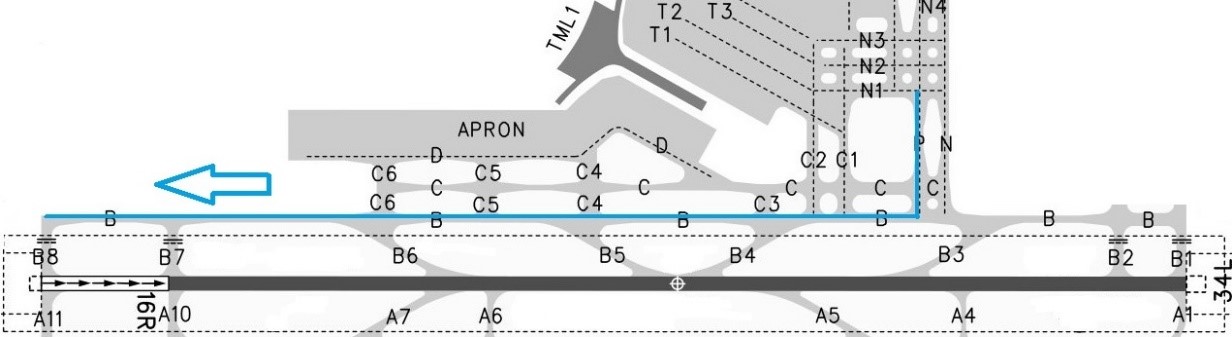}
\caption{The specific taxiing path of aircraft in Command B}
\label{fig8}
\end{figure}

Figure 9 shows the deduction effect of the aircraft taxiing process in Command B.

\begin{figure}[ht]
%\centering\includegraphics[width=2.5in]{xxxxxx.eps}
%%% where xxxxxx name represents "figurename.eps"
\centering\includegraphics[width=4.5in]{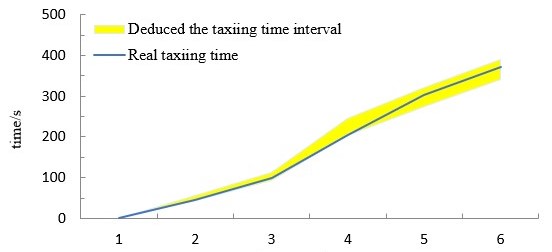}
\caption{The effect diagram of the deductive taxiing process of aircraft in Command B}
\label{fig9}
\end{figure}

In the same way, another real control routing command in the target airport is selected and recorded as Command C. By querying the established information table of all operating aircraft belonging to the target airport, the aircraft type is B737-800, and the aircraft registration number is B-7893. The taxiways involved in the aircraft taxiing in the control command are C1, B-4, B-6, and B-8 in sequence. When the controller issued the command, it was daytime on a weekday, so each segment of the taxi-way that the aircraft passed through called its own $\Delta$$Vk_{wdd}$ as the taxiing speed interval. The specific taxiing path of the aircraft is shown in figure 10.

\begin{figure}[ht]
%\centering\includegraphics[width=2.5in]{xxxxxx.eps}
%%% where xxxxxx name represents "figurename.eps"
\centering\includegraphics[width=5in]{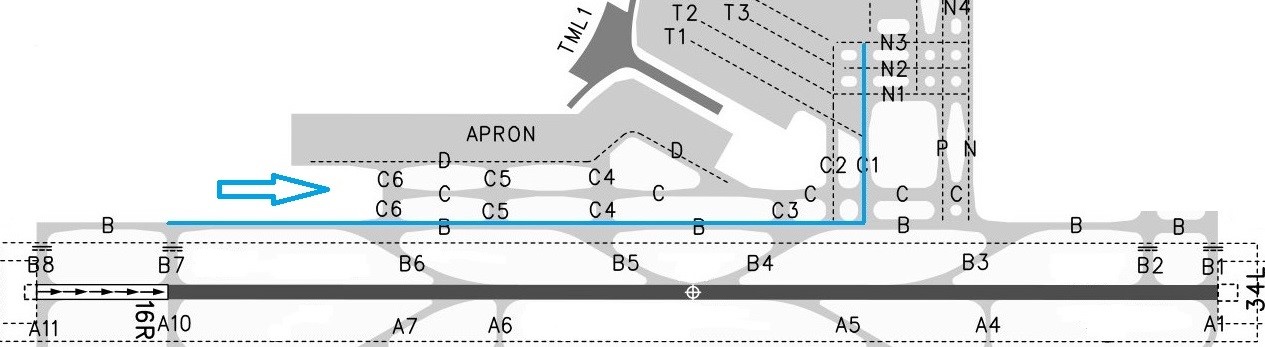}
\caption{The specific taxiing path of aircraft in a Command C}
\label{fig10}
\end{figure}

The deduction effect of the aircraft taxiing process in Command C is shown in figure 11.

\begin{figure}[ht]
%\centering\includegraphics[width=2.5in]{xxxxxx.eps}
%%% where xxxxxx name represents "figurename.eps"
\centering\includegraphics[width=4.5in]{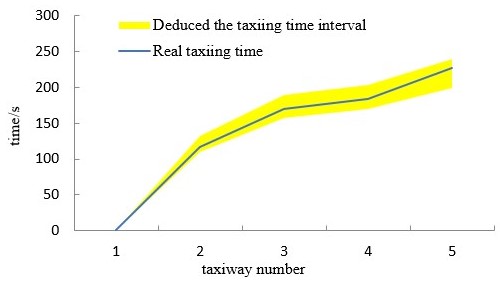}
\caption{The effect diagram of the deductive taxiing process of aircraft in Command C}
\label{fig11}
\end{figure}

From the above, the average deviation of the three deduced taxiing time intervals from the real taxiing time and the correlation coefficient are shown in table 7.

%下方文字到表格上方 另起一夜
\clearpage

\begin{table}[ht]
\caption{The average deviation and correlation coefficient between the three deduced taxi time intervals and the real taxi time}%%%Table caption goes here
\label{table7}
\centering
 \begin{tabular}{m{2.3cm}<{\centering}m{2.3cm}<{\centering}m{2.3cm}<{\centering}m{2.6cm}<{\centering}m{2.6cm}<{\centering}}%%%The number of columns has to be defined here
\hline
Command&Average deviation /\%  &Correlation coefficient&The maximum value of the deduction time interval/$s$ &The average length of the deduced interval/$s$ \\
\hline
   Command A &-2.25 &0.98 &38.6 &23.6 \\
   Command B &3.69 &0.97 &49.1 &34.1\\
   Command C &0.78 &0.99 &40.3 &31.8 \\
    \hline
\end{tabular}
\end{table}%%%End of the table

Through the analysis of the results in table 7, it can be seen that the real taxiing time of the aircraft in the three command is completely included in the time interval obtained by the deduction. The average deviations of the three deduced taxiing time intervals from the real time are all close to 0. The correlation coefficients between the three obtained deduce taxiing time intervals and the real taxiing time are all close to 1.

In the same way, a total of 50 groups of 50 real control routing command in the target airport are selected to conduct a simulation experiment to deduce the taxiing process of the aircraft, and the average deviation between the deduced taxiing time interval and the real taxiing time is 3.14\%. The mean correlation coefficient over time was 0.97.

Through the analysis of the above results, it can be seen that under the deduction method of the aircraft taxiing process proposed in this chapter, the correlation coefficients between the deduced taxiing time interval and the real taxiing time are all close to 1, indicating that the correlation is strong and the degree of correlation is good, which is a complete correlation. At the same time, the average deviation between the obtained deduction taxiing time interval and the real taxiing time is close to 0, indicating that the two have good consistency and high degree of fitting, and can better fit the changes of the real value, which is closer to the real situation of the aircraft taxiing process.

\subsubsection{Simulation and comparison experiments to verify and deduce the taxiing process of the aircraft}

The same real control command from the target airport is selected as the simulation comparison experiment to verify the effect of the taxiing process of the deduced aircraft. Select the real control command C in the previous section, and compare the effect of deduced the taxiing process of the aircraft through the method of deduced the taxiing process based on the taxiing speed interval established under different constraints.

First, SU et al.\cite{ref29} were used to deduce the taxiing process of the aircraft, in which the speed on the straight taxiway was defined as 0 to 30 knots, and the speed on the apron taxiway was defined as 0 knots to 10 knots. The acceleration or deceleration process was considered as a uniform acceleration (or deceleration) movement, and the acceleration is defined as $\pm$\SI{2}{\meter\per\second^2}, which is recorded as the uniform acceleration and deceleration deduction method. figure 12 shows the deduction effect of the aircraft taxiing process of the uniform acceleration and deceleration deduction method.

%下方文字到表格上方 另起一夜
\clearpage

\begin{figure}[ht]
%\centering\includegraphics[width=2.5in]{xxxxxx.eps}
%%% where xxxxxx name represents "figurename.eps"
\centering\includegraphics[width=4.5in]{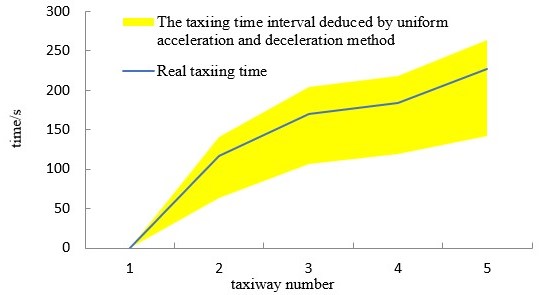}
\caption{The effect diagram of aircraft taxiing process deduced by the uniform acceleration and deceleration method}
\label{fig12}
\end{figure}

Secondly, in order to compare the method of deducting the taxiing process of defining the taxiing speed interval based on historical data, the taxiing speed interval of the aircraft on each taxiway is not divided into four parts according to the taxiing time, and only the taxiing after the Pauta criterion is used. The method of deriving the taxiing process in the speed range $\Delta$$Vk-Pauta$) is recorded as the $\Delta$$Vk-Pauta$ deduction method. The effect of deduce the taxiing process of the aircraft is shown in figure 13.

\begin{figure}[ht]
%\centering\includegraphics[width=2.5in]{xxxxxx.eps}
%%% where xxxxxx name represents "figurename.eps"
\centering\includegraphics[width=4.5in]{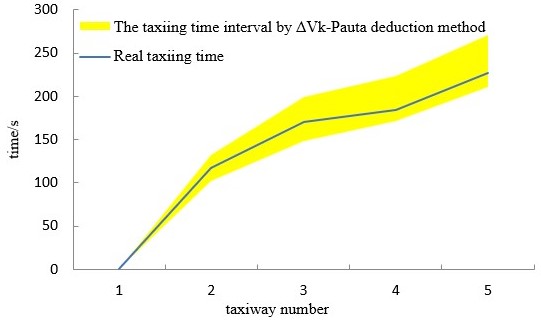}
\caption{ The effect diagram of aircraft taxiing process deduced by the $\Delta$$Vk-Pauta$ method}
\label{fig13}
\end{figure}

Finally, the deduction is carried out according to the deduction method of the aircraft taxiing process proposed in this chapter. The method has been verified in the previous section and will not be repeated here.

The average deviation and correlation coefficient between the deduced taxiing time interval and the real taxiing time obtained by the above three methods are shown in table 8.

\begin{table}[h!]
\caption{The average deviation and correlation coefficient between the deduced taxi time interval and the real taxi time obtained by the three methods}%%%Table caption goes here
\label{table8}
\centering
 \begin{tabular}{m{2.9cm}<{\centering}m{2cm}<{\centering}m{2cm}<{\centering}m{2.6cm}<{\centering}m{2.6cm}<{\centering}}%%%The number of columns has to be defined here
\hline
Method&Average deviation /\%  &Correlation coefficient&The maximum value of the deduction time interval/$s$ &The average length of the deduced interval/$s$ \\
\hline
 Uniform acceleration and deceleration deduction method &20.59 &0.79 &121.7 &99.2 \\
 \\
 $\Delta$$Vk-Pauta$ deduction method&9.27 &0.90 &59.6&48.3\\
 \\
 Taxiing process deduction method &0.78 &0.99 &40.3 &31.8 \\
    \hline
\end{tabular}
\end{table}%%%End of the table

Through the analysis of the results in table 8, it can be seen that under the same control command, the real taxiing time of the three methods is completely included in the time interval obtained by the deduction. However, the amplitude and average length of the deduced taxiing time interval obtained by the aircraft taxiing process deduction method proposed in this chapter are significantly smaller than those obtained by the uniform acceleration and deceleration method and the  $\Delta$$Vk-Pauta$ deduction method, which shows the time margin of the aircraft taxiing process deduction method proposed in this chapter. Smaller, indicating that the deduction is more accurate and reasonable. And the correlation coefficient between the deduced taxiing time interval and the real taxiing time is closer to 1 than that of the uniform acceleration and deceleration deduction method and the  $\Delta$$Vk-Pauta$ deduction method, indicating that the correlation is stronger and the degree of correlation is better, which is a complete correlation. In addition, the average deviation between the deduced taxiing time interval and the real time is closer to 0 than the uniform acceleration and deceleration deduction method and the  $\Delta$$Vk-Pauta$ deduction method, indicating that the two have better consistency and higher fitting degree, and can better fit the changes of the real value. The above results show that the deduction method of the aircraft taxiing process proposed in this chapter has a better deduction effect and is closer to the real situation of the aircraft taxiing process. In addition, due to the influence of weather, road conditions and other factors on the taxiing of the aircraft, there must be certain errors in the aircraft taxiing deduction. However, according to the experimental results of the aircraft taxiing deduction, the effect of the error on the taxiing process of the aircraft is within a reasonable range. To sum up, it is believed that the aircraft taxiing process derivation method based on the control command proposed in this paper can accurately deduce the future aircraft taxiing process.

\section{ Aircraft Taxiing Conflict Detection and Early Warning Method}

In recent years, with the increasing congestion in the airfield of the airport, it is not uncommon for aircraft to have conflict accidents and incidents while taxiing in the airfield. Therefore, how to solve and deal with the problem of taxiing conflicts, how to improve the special situation handling ability and emergency early warning ability of the airport flight area, and how to provide timely decision-making support and effective assistance to the controller through technical means are of great importance to improve the safety of airport operation.

Since there is currently no way to verify the control command immediately after the controller has issued the command, another purpose of this paper is to provide a way to verify the control order in advance before the aircraft starts to actually taxi out/in and after the controller has issued the order. Correctness and solutions to achieve aircraft taxiing conflict detection and early warning.

Aiming at the above problems, this paper proposes a new paradigm of aircraft taxiing conflict detection and early warning method. In the research based on the deduction of the taxiing process of the aircraft in the previous part, the taxiing speed of the aircraft on the taxiway is also defined as different interval variables. After the explanation in the previous part, this fitting process is closer to the real taxiing process of the aircraft.

\subsection{Aircraft Taxiing Conflict Detection Algorithm}

Before designing the aircraft taxiing conflict detection algorithm, the minimum safety time threshold between aircraft ($T_{no}$) is set. In reality, there are only two states of aircraft taxiing conflict: no conflict and conflict. $T_{no}$ is used as the conflict threshold to determine whether the aircraft is completely conflict-free or out of conflict.$T_{no}$ consists of three parts: the minimum time thresh-old of conflict aircraft $T_c$), the braking coordination time of aircraft ($T_{bc}$) and the pilot's reaction time ($T_r$). According to NASA's research on aircraft collision parameters on the airport surface, the pilot's reaction time ($T_r$) is 2 seconds. In a related study, $T_{bc}$ was shown to be 1.2 seconds. Tc is divided into two categories according to the facing angle between the planes, one is the mini-mum time threshold of the aircraft in the head-to-head conflict ($T_{c1}$), and the other is the minimum time threshold of the other types of conflict aircraft ($T_{c2}$). When there is a head-to-head conflict between the aircraft, their taxiing directions are opposite. According to the rules, the maximum taxiing speed ($V_{max}$) of the aircraft when taxiing on ground cannot exceed \SI{50}{km/h} and the minimum distance ($L_{min}$) between the aircraft when taxiing on ground cannot be less than \SI{50}{m} two conditions, to calculate $T_{c1}$.

\begin{align}\label{4.1}
\begin{split}
T_{c 1}=\frac{L_{\text {min }}}{V_{\max }+V_{\max }}
\end{split}
\end{align}

According to formula (4.1), it can be calculated that $T_{c1}$ is 1.8 seconds. Similarly, calculate the minimum time threshold $T_{c2}$ for other types of conflicting aircraft.

\begin{align}\label{4.2}
\begin{split}
T_{c 2}=\frac{L_{\min }}{V_{\max }}
\end{split}
\end{align}

According to formula (4.2), it can be calculated that $T_{c2}$ is 3.6 seconds.

To sum up, the minimum safe time threshold $T_{no}$ between aircraft is divided into two categories according to the type of conflict. One is the minimum time safety threshold of the aircraft in the head-to-head conflict ($T_{no1}$).

\begin{align}\label{4.3}
\begin{split}
T_{n o 1}=T_{c 1}+T_{b c}+T_{r}
\end{split}
\end{align}

According to formula (4.3), it can be calculated that $T_{n o 1}$ is 5 seconds. The other is the minimum time safety threshold for other types of conflict aircraft ($T_{n o 2}$).

\begin{align}\label{4.4}
\begin{split}
T_{n o 2}=T_{c 2}+T_{b c}+T_{r}
\end{split}
\end{align}

According to formula (4.4), it can be calculated that $T_{n o 2}$ is 6.8 seconds.
After obtaining the minimum safe time threshold between aircrafts ($T_{n o}$), the designed algorithm flow of taxiing conflict detection is shown in figure 14. The specific steps are as follows.

Step 1: Enter the taxiway number $k$ in the two routing command $P_i$ and $P_j$ and the start taxiing times $t_{s i}$ and $t_{s j}$ of the aircraft $P_i$ and $P_j$, where $P_i$ and $P_j$ represent sets of different taxiing aircraft.

Step 2: Traverse the intersection nodes on all taxiing paths between aircraft. From the deduction taxiing time interval for air-craft $P_i$ to reach taxiway $k$ ($\Delta tp_{i}$) and the deduction taxiing time interval for aircraft $P_j$ to arrive at taxiway $k$ ($\Delta tp_{j}$) calculated in the previous part, and then calculate the taxiing time interval difference between aircraft $P_i$ and $P_j$ passing through the same common node or taxiway value ($\Delta tp_{i j}$).

\begin{align}\label{4.5}
\begin{split}
\Delta t P_{i j}=\Delta t P_{i}-\Delta t P_{j}
\end{split}
\end{align}

The lower limit of $\Delta tp_{i j}$ is defined as $ t_{min}p_{ij}$, and the upper limit of $\Delta tp_{i j}$ is defined as$ t_{max}p_{i j}$.

Step 3: Determine whether the heading angle between the planes is $180^{\circ}$, if so, call $T_{n o 1}$; if not, call $T_{n o 2}$.

Step 4: Perform conflict detection: $T_{n o 1}$ and $T_{no2}$ both belong to $T_{n o}$. Next, for the convenience of expression, $T_{n o}$ represents $T_{n o 1}$ and $T_{n o 2}$ for judgment. If $ t_{min}p_{i j} \leqslant T_{n o} \leqslant  t_{max}p_{i j}$, then the conflict probability $p_{i j}$.

\begin{align}\label{4.6}
\begin{split}
p_{i j}=\frac{T_{n o}-t_{\min } p_{i j}}{t_{\max } p_{i j}-t_{\min } p_{i j}} * 100 \%
\end{split}
\end{align}

Step 5: If $T_{no} <  t_{min}p_{ij}$, traverse the intersection nodes on all taxiing paths of each aircraft again. If $T_{no} >  t_{max}p_{ij}$, the collision probability is recorded as $100\%$.

\subsection{Classification of Warning Levels}

Early warning refers to the prediction and forecasting of crisis events before a crisis occurs. It first appeared in military fields such as early warning radar, and then gradually expanded to various fields such as transportation, meteorology, economy, and medical care, such as fatigue driving early warning systems, earthquake early warning systems, financial early warning systems, and medical security risk early warning systems. Applied to the detection of aircraft taxiing conflicts, early warning is to predict potential conflicts in advance and send out alarm information during the taxiing process of aircraft, so that controllers can adjust control command in advance to avoid accidents and incidents.

The early warning level is divided according to the risk degree, emergency degree, development trend, possible harm degree and influence scope of accidents and incidents. According to Article 14 of China's "Administrative Measures for Civil Aviation Air Traffic Management Safety Assessment", the event risk analysis should use qualitative or quantitative methods to start from the severity of the consequences and the possibility of occurrence, comprehensively evaluate the size of the risk, and determine whether it is acceptable or not based on the risk acceptability criteria. table 9 and table 10 show the classification table of the severity of risk consequences and the possibility of risk occurrence.

\begin{table}[ht]
\caption{ Classification of severity of dangerous consequences}%%%Table caption goes here
\label{table9}
\centering
 \begin{tabular}{m{1.6cm}<{\centering}m{2cm}<{\centering}m{2cm}<{\centering}m{2cm}<{\centering}m{2cm}<{\centering}m{2cm}<{\centering}}%%%The number of columns has to be defined here
\hline
 &Ignorable&Slight&Serious&Dangerous&Catastrophic \\
\hline
Qualitative description &ATC service level is a little bit reduced, or the spacing is a little bit reduced &ATC service level is slightly reduced, or the spacing is slightly reduced &ATC is unable to provide some services, or the distance loss is large &ATC unable to provide service, or loss of safety distance& The aircraft collided or collided with the ground or obstacles\\
    \hline
\end{tabular}
\end{table}%%%End of the table

\begin{table}[ht]
\caption{ Classification of the probability of occurrence of the hazard}%%%Table caption goes here
\label{table10}
\centering
 \begin{tabular}{m{1.7cm}<{\centering}m{2cm}<{\centering}m{2cm}<{\centering}m{2cm}<{\centering}m{2cm}<{\centering}m{2cm}<{\centering}}%%%The number of columns has to be defined here
\hline
 &Highly unlikely &Rare &Accidental &Regular &Frequent \\
\hline
Quantitative description (per operating hour) &$<1 \times 10^{-9}$& $\ge 1 \times 10^{-9}<1 \times 10^{-7} $& $\ge 1 \times 10^{-7}<1 \times 10^{-5} $&$\ge1 \times 10^{-5}< 1 \times 10^{-3}$& $\ge 1 \times 10^{-3}$\\
\\
Qualitative description (single unit)&Not once in 100 years&Occurs once every $10\thicksim 100$ years &Occurs approximately once a year&Occurs approximately once a month&Occurs more than once a week\\
\\
Qualitative description (system-wide)&Not once in 30 years&Occurs once every 3 years&Occurs once every few months&Occurs several times a month&Occurs once every $1\thicksim 2$ days\\
    \hline
\end{tabular}
\end{table}%%%End of the table

Risk acceptability is an important basis for risk analysis, and risks are divided into unacceptable risks, tolerable risks and acceptable risks according to their acceptability. Corresponding to the research field of aircraft taxiing conflict warning, refer to table 9 and table 10. According to whether the detected conflict probability will affect the airport operation and the controller's scheduling control work after the aircraft is disturbed, the aircraft taxiing conflict early warning is divided into four levels: low, medium, high and dangerous. Setting the aircraft taxiing conflict early warning mechanism to multi-level early warning feedback, multi-level early warning is more humane, and can provide controllers with various ways to take measures.

(1) Low-level early warning

When the service level of Air Traffic Control (ATC) is a little bit reduced or the taxiing spacing of the aircraft is a little bit reduced, this alarm situation is classified as a low-level warning, and the alarm situation is minor, which corresponds to an acceptable risk. At this time, the controller does not need to adjust the control command that has been issued, and does not need to modify the issued command route.

(2) Intermediate-level early warning

When the ATC service level is slightly reduced or the taxiing interval of the aircraft is slightly reduced, the aircraft taxiing routing plan may need to be adjusted. At this time, the controller needs to follow up on the taxiing dynamics of the aircraft, and modify the issued command route if necessary. After the adjustment, the controller must ensure that other aircraft can operate normally, and the aircraft interfered by the conflict will not cause secondary conflicts due to the adjustment of the control routing plan. At the same time, the controller can remind the pilot where a conflict may occur.

(3) High-level warning

When ATC is unable to provide some services or the taxiing interval of the aircraft is lost greatly. In this case, the aircraft taxiing route plan needs to be readjusted, and the workload of control, dispatch and command will increase. Therefore, this kind of alarm situation is classified as high-level early warning. The alarm situation is more serious and corresponds to an unacceptable risk. At this time, the controller needs to focus on this area, and must modify the issued control command according to the current operating conditions of the airport flight area. Similarly, the controller must ensure that other aircraft can operate normally after adjustment.

(4) Dangerous warning

When ATC cannot provide service or the taxiing safety distance of the aircraft is completely lost, that is, the conflict probability mentioned in the aircraft taxiing conflict detection algorithm in the previous section reaches $100\%$. In this case, the aircraft taxiing routing plan needs to be readjusted. The workload of the controller's control, dispatch and command will definitely in-crease, so this kind of warning is classified as a danger warning. The warning is very serious and corresponds to an unacceptable risk. At this point, it means that the taxiing time difference between planes has fallen below the minimum time safety threshold between planes, and the controller must immediately modify the issued control command according to the current operating conditions of the airport flight area.

The maximum membership principle in fuzzy mathematics is adopted for how to judge which early warning level the taxiing conflict probability between aircrafts detected by the aircraft taxiing conflict detection algorithm belongs. The principle is defined as follows.

Suppose $m$ fuzzy subsets $A_1, A_2, …, A_m$, on the universe $U=\left\{x_{1}, x_{2}, \ldots, x_{n}\right\} $, if for any $x_{0} \in U$, there is $i_{0} \in\{1,2, \cdots, m\}$ , such that

\begin{align}\label{4.7}
\begin{split}
A_{i_{0}}\left(x_{0}\right)=\bigvee_{k=1}^{m} A_{k}\left(x_{0}\right)
\end{split}
\end{align}

Then it is considered that $x_{0}$ is relatively affiliated to $A_{i_{0}}$ .

The fuzzy membership function formula for the classification of aircraft taxiing conflict early warning levels is as follows.

\begin{align}\label{4.8}
\begin{split}
A(x)=\left\{\begin{array}{cl}
1, & x<a \\
\left(\frac{b-x}{b-a}\right)^{2}, & a \leqslant x \leqslant b \\
0, & b<x
\end{array}\right.
\end{split}
\end{align}

When setting the threshold parameter of conflict warning, it should not be too large, which will cause too many false alarms, nor should it be too small, which will cause poor warning effect. According to the feedback statistics of multiple aircraft taxiing conflict detection experiments, the conflict probability thresholds between the four early warning levels are set as 0, 0.32, 0.61 and 1, respectively. The setting of the conflict probability threshold is not static, it needs to be determined according to the operating parameters of the aircraft taxiing conflict detection experiment in different airports.

Fuzzy membership function of low-level warning is as follows.

\begin{align}\label{4.9}
\begin{split}
A(x)=\left\{\begin{array}{cc}
1, & x<0.32 \\
\left(\frac{0.61-x}{0.29}\right)^{2}, & 0.32 \leqslant x \leqslant 0.61 \\
0, & 0.61<x
\end{array}\right.
\end{split}
\end{align}

Fuzzy membership function of intermediate-level early warning is as follows.

\begin{align}\label{4.10}
\begin{split}
A(x)=\left\{\begin{array}{lc}
\left(\frac{x}{0.32}\right)^{2}, & 0<x<0.32 \\
1, & 0.32<x<0.61 \\
\left(\frac{1-x}{0.39}\right)^{2}, & 0.61 \leqslant x \leqslant 1 \\
0, & x<0 \text { or } 1<x
\end{array}\right.
\end{split}
\end{align}

Fuzzy membership function for advanced warning is as follows.

\begin{align}\label{4.11}
\begin{split}
A(x)=\left\{\begin{array}{cc}
0, & x<0.32 \\
\left(\frac{x-0.32}{0.29}\right)^{2}, & 0.32 \leqslant x \leqslant 0.61 \\
1, & 0.61<x
\end{array}\right.
\end{split}
\end{align}

For the situation where the danger warning level corresponds to the aircraft taxiing conflict probability detected in the aircraft taxiing conflict detection algorithm in Section 4.1 is $100\%$.

Combining the above proposed aircraft taxiing conflict detection algorithm and the proposed method of classifying early warning levels, a deduction-based aircraft taxiing conflict detection and early warning method is obtained. The specific process of aircraft taxiing conflict detection and early warning method is shown in figure 14.
\begin{figure}[ht]
%\centering\includegraphics[width=2.5in]{xxxxxx.eps}
%%% where xxxxxx name represents "figurename.eps"
\centering\includegraphics[width=5in]{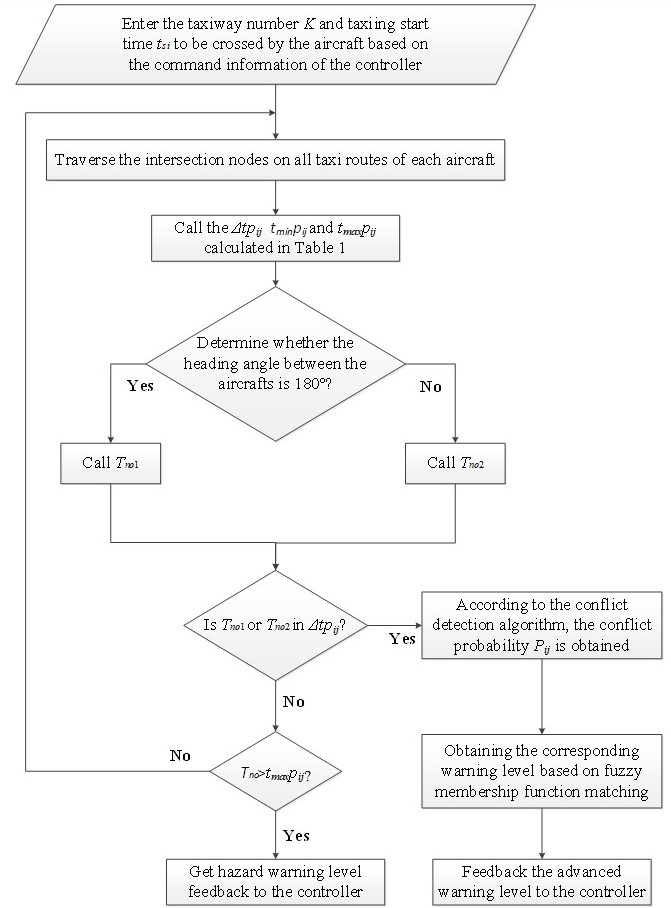}\caption{ Flow chart of aircraft taxiing conflict detection and early warning method}
\label{fig14}
\end{figure}

\subsection{ Simulation Experiment and Comparative Experiment Verification of Aircraft Taxiing Conflict Detection and Early Warning Method }

Firstly, the method proposed in this chapter is used to conduct a simulation experiment of aircraft taxiing conflict detection and early warning on multiple sets of real data in the target airport. The results show that there is no corresponding level of conflict early warning. Through the verification of the target airport accident and incident reports in the same period, there were no aircraft taxiing conflict accidents and incidents in the same period, which preliminarily shows the effectiveness of the method proposed in this paper. However, no taxiing conflict accidents and incidents occurred at the target airport in that period, which does not mean that taxiing conflicts did not occur between aircrafts. According to the actual airport situation, when the aircraft is near the taxiway intersection, it may encounter other aircraft, and there may be cross or confrontation conflicts. Incoming aircraft may encounter confrontation conflicts with departing aircraft when they are about to enter the apron system.

Therefore, in order to verify the effect of aircraft taxiing conflict detection and early warning again, the idea of dynamic taxiing time window is used for reference, and different real aircraft taxiing processes are translated on the time axis to obtain data of different types of taxiing conflicts. Although the start time of the aircraft taxiing is different, the taxiing process of the aircraft is real.
When a confrontation conflict occurs with the aircraft, the minimum time safety threshold $T_{n o 1}$ of the aircraft in the confrontation conflict is called. When cross and rear-end conflict occur with the aircraft, the minimum time safety threshold $T_{n o 2}$ for other types of conflict aircraft is called. The time difference between the aircrafts in the real command is defined as $t_{i j}$, and the two conflict situations when the real aircraft is taxiing are divided as follows: when $T_{n a} \geq t_{i j}$, it means that there is a conflict between the aircraft; when $T_{n o} < t_{i j}$, it means that there is no conflict between the aircraft.

\subsubsection{Simulation Experiment Verification of Aircraft Taxiing Confrontation Conflict}

First of all, in view of the taxiing conflict between the planes, two real control routing command with 24-bit aircraft address codes 78142A and 780FDF in the target airport are selected. The route, speed and total time of the whole taxiing process of the two aircraft are not changed, and only the taxiing start time are adjusted. The real taxiing paths of the two aircraft are shown in figure 15.

\begin{figure}[ht]
%\centering\includegraphics[width=2.5in]{xxxxxx.eps}
%%% where xxxxxx name represents "figurename.eps"
\centering\includegraphics[width=5in]{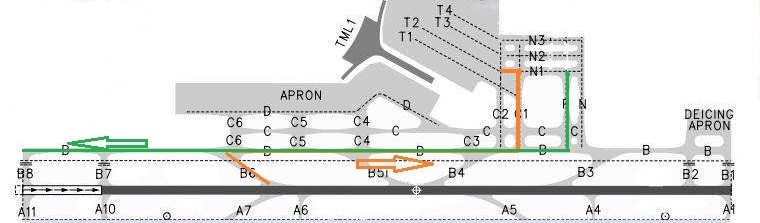}
\caption{  The real taxiing path diagram of 78142A and 780FDF}
\label{fig15}
\end{figure}

It can be clearly seen from figure 15 that the area where a confrontation conflict may occur is taxiway B (area C3 to C6).  Considering that during the actual operation of the target airport, the time interval for the controller to arrange the aircraft to start taxiing is generally within the range of one and a half minutes, so the maximum value of the difference between the taxiing time of the aircraft is set to \SI{100}{seconds}. According to the data characteristics of the time difference between the start of taxiing of the aircraft, it is observed that the conflict detection results are relatively stable within the range of \SI{5}{seconds}. Therefore, the difference in the starting taxiing time of the aircraft is divided into 21 groups from \SI{0}{seconds} to \SI{100}{seconds} in units of \SI{5}{seconds}. The specific results of the experiment to verify this confrontation conflict are shown in table 11.

\begin{table}[ht]
\caption{Authenticate confrontation conflict}%%%Table caption goes here
\label{table11}
\centering
 \begin{tabular}{m{2.5cm}<{\centering}m{2.5cm}<{\centering}m{2.5cm}<{\centering}m{4cm}<{\centering}}%%%The number of columns has to be defined here
\hline
Offset of aircraft taxiing start time/s &Real conflict situation&Conflict probability/\%&Experimental warning level \\
 \hline
0 &No conflict &32.90 &Intermediate-level warning \\
5 &No conflict &40.09 &Intermediate-level warning \\
10 &No conflict &47.28 &Intermediate-level warning \\
15 &No conflict &47.28 &Intermediate-level warning \\
20 &No conflict &61.66 &Intermediate-level warning \\
25 &No conflict &50.71 &Intermediate-level warning \\
30 &No conflict &50.24 &Intermediate-level warning \\
35 &No conflict &57.79 &Intermediate-level warning \\
40 &No conflict &65.33 &Intermediate-level warning \\
45 &No conflict &47.64 &Intermediate-level warning \\
50 &No conflict &40.10 &Intermediate-level warning \\
55 &No conflict &32.55 &Intermediate-level warning \\
60 &No conflict &25.01 &Low-level warning \\
65 &No conflict &17.46 &Low-level warning \\
70 &No conflict &22.92 &Low-level warning \\
75 &No conflict &31.82 &Intermediate-level warning \\
80 &No conflict &40.71 &Intermediate-level warning \\
85 &No conflict &49.61 &Intermediate-level warning \\
90 &No conflict &58.51 &Intermediate-level warning \\
95 & conflict &67.40 &High-level warning \\
100 & conflict &76.30 &High-level warning \\
    \hline
\end{tabular}
\end{table}%%%End of the table

In the process of verifying the conflict detection of the taxiing conflict between the two aircrafts, the taxiing trajectories and processes of the two aircrafts are real and unique, so the real conflict situation is also determined. In table 11, the two real conflict situations (conflict and non-conflict) when the aircraft are taxiing are obtained by comparing the time difference between judging the real command aircraft and the minimum safe time threshold between aircraft defined in Section 4.1. The conflict probability and the experimental warning level are calculated by the aircraft taxiing conflict detection algorithm proposed in Sections 4.1 and the method of dividing the warning level proposed in Sections 4.2. The experimental early warning level is divided according to the early warning level in Section 4.2, among which, acceptable risk corresponds to low-level early warning in aircraft taxiing conflict detection and early warning method, tolerable risk corresponds to intermediate-level early warning, and unacceptable risk corresponds to high-level and dangerous early warning.

It can be seen from table 11 that when the difference between the start taxiing time is between \SI{95}{s} and \SI{100}{s}, the experimental early warning level is high-level early warning. After the real conflict situation test, the opponent conflict is true, and there is no false alarm or missed alarm. The experimental results show that the real confrontation conflict situation in the possible confrontation conflict area of two aircraft is consistent with the experimental conflict warning level, which shows that the proposed method can effectively realize the simulation experiment verification of confrontation taxiing conflict detection and early warning of aircraft based on real control command conditions.

\subsubsection{Simulation Experiment Verification of Aircraft Taxiing Cross Conflict}

Similarly, for aircraft taxiing cross conflict, the 24-bit aircraft address code in the target airport is selected as two real control routing command, 7808B7 and 780FDF. The real taxiing paths of the two aircraft are shown in figure 16.

\begin{figure}[ht]
%\centering\includegraphics[width=2.5in]{xxxxxx.eps}
%%% where xxxxxx name represents "figurename.eps"
\centering\includegraphics[width=5in]{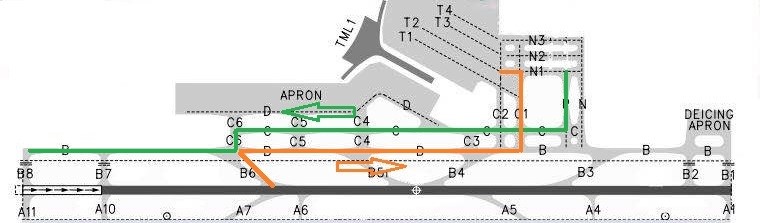}
\caption{The real taxiing path diagram of 7808B7 and 780FDF}
\label{fig16}
\end{figure}

It can be clearly seen from figure 16 that the point where the two aircraft may have a conflict is the intersection of taxiway C and taxiway C1. In the same way, the deduced taxiing processes of two aircraft are compared with the real taxiing processes. The results of verifying this cross conflict are shown in table 12.

\begin{table}[ht]
\caption{Authenticate cross conflict}%%%Table caption goes here
\label{table12}
\centering
 \begin{tabular}{m{2.5cm}<{\centering}m{2.5cm}<{\centering}m{2.5cm}<{\centering}m{4.5cm}<{\centering}}%%%The number of columns has to be defined here
\hline
Offset of aircraft taxiing start time/s &Real conflict situation&Conflict probability/\%&Experimental warning level \\
 \hline
0 &No conflict &9.43 &Low-level warning \\
5 &No conflict &18.29 &Low-level warning \\
10 &No conflict &27.84 &Low-level warning \\
15 &No conflict &36.01 &Intermediate-level warning \\
20 &No conflict &44.87 &Intermediate-level warning \\
25 &No conflict &53.73 &Intermediate-level warning \\
30 & conflict &61.51 &High-level warning \\
35 &No conflict &52.65 &Intermediate-level warning \\
40 &No conflict &43.78 &Intermediate-level warning \\
45 &No conflict &34.92 &Intermediate-level warning \\
50 &No conflict &26.06 &Low-level warning \\
55 &No conflict &17.20 &Low-level warning \\
60 &No conflict &8.34 &Low-level warning \\
65 &No conflict &0 &Low-level warning \\
70 &No conflict &0 &Low-level warning \\
75 &No conflict &0 &Low-level warning \\
80 &No conflict &0 &Low-level warning \\
85 &No conflict &0 &Low-level warning \\
90 &No conflict &0 &Low-level warning \\
95 &No conflict &0 &Low-level warning \\
100 &No conflict &0 &Low-level warning \\
    \hline
\end{tabular}
\end{table}%%%End of the table

In the experimental verification process of detecting aircraft cross conflict, the verification principle is consistent with Section 4.3.1. It can be seen from table 12 that when the time difference between the start of taxiing is \SI{30}{s}, the early warning level of the experiment is high-level early warning. After the test of the real conflict situation, the cross conflict is true, and there are no false alarms and missed alarms. The experimental results show that the real cross conflict situation in the possible cross conflict area of two aircraft is consistent with the experimental conflict warning level, which shows that the proposed method can effectively realize the simulation experiment verification of cross taxiing conflict detection and early warning of aircraft based on real control command conditions.

\subsubsection{ Simulation Experiment Verification of Aircraft Rear-end Conflict}

In the same way, for the conflict of aircraft taxiing and rear-end conflict, the 24-bit aircraft address code in the target airport is selected as two real control routing command, 780D33 and 78142A. The real taxiing paths of the two aircrafts are shown in figure 17.

\begin{figure}[ht]
\centering\includegraphics[width=5in]{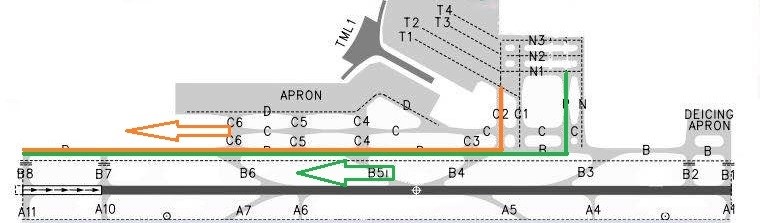}
\caption{ The real taxiing path diagram of 780D33 and 78142A}
\label{fig17}
\end{figure}

It can be clearly seen from the figure 17 that the area where the two aircraft may rear-end is taxiway B (area C2 to B8). Similarly, the deduced taxiing processes of two aircraft are compared with the real taxiing processes. The results of the rear-end conflict are shown in table 13.

In the experimental verification process of detecting aircraft rear-end conflict, the verification principle is the same as that in Section 4.3.1. It can be seen from table 13 that when the time difference between the start of taxiing is \SI{35}{s}, the experimental early warning level is high-level early warning. After the real conflict situation test, the rear-end collision is true, and there is no false alarm or missed alarm. The experimental results show that the real rear-end conflict situation in the possible rear-end conflict area of two aircraft is consistent with the experimental conflict warning level, which shows that the proposed method can effectively realize the simulation experiment verification of rear-end taxiing conflict detection and early warning of aircraft based on real control command conditions.

\clearpage

\begin{table}[ht]
\caption{ Authenticate rear-end conflict}%%%Table caption goes here
\label{table13}
\centering
 \begin{tabular}{m{2.5cm}<{\centering}m{2.5cm}<{\centering}m{2.5cm}<{\centering}m{4.5cm}<{\centering}}%%%The number of columns has to be defined here
\hline
Offset of aircraft taxiing start time/s&Real conflict situation&Conflict probability/\%&Experimental warning level \\
 \hline
0 &No conflict &16.54 &Low-level warning \\
5 &No conflict &22.19 &Low-level warning \\
10 &No conflict &27.84 &Low-level warning \\
15 &No conflict &33.49 &Intermediate-level warning \\
20 &No conflict &39.15 &Intermediate-level warning \\
25 &No conflict &44.80 &Intermediate-level warning \\
30 &No conflict &52.45 &High-level warning \\
35 & conflict &62.50 &Intermediate-level warning \\
40 &No conflict &53.62 &Intermediate-level warning \\
45 &No conflict &42.31 &Intermediate-level warning \\
50 &No conflict &36.66 &Low-level warning \\
55 &No conflict &31.04 &Low-level warning \\
60 &No conflict &25.35 &Low-level warning \\
65 &No conflict &19.70 &Low-level warning \\
70 &No conflict &14.04 &Low-level warning \\
75 &No conflict &8.39 &Low-level warning \\
80 &No conflict &2.74 &Low-level warning \\
85 &No conflict &0 &Low-level warning \\
90 &No conflict &0 &Low-level warning \\
95 &No conflict &0 &Low-level warning \\
100 &No conflict &0 &Low-level warning \\
    \hline
\end{tabular}
\end{table}%%%End of the table

To sum up, the experimental verification is carried out with 50 sets of flight operation data in the target airport on September 1, 2019. For the three conflicts, the aircraft taxiing conflict detection and early warning method based on the $\Delta Vk-time$ taxiing speed interval is also used for simulation experiments. Figure 18 shows the accuracy of the three conflict warning levels obtained.

\begin{figure}[ht]
\centering\includegraphics[width=5in]{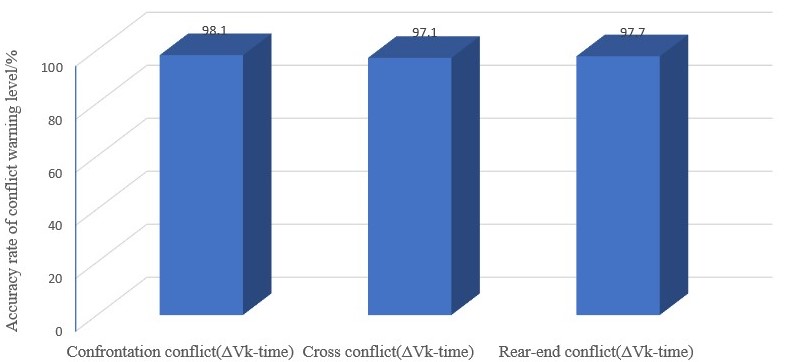}
\caption{The accuracy of the three conflict warning levels obtained}
\label{fig18}
\end{figure}

Through the analysis of the results in figure 18, it can be seen that the aircraft taxiing conflict detection and early warning method based on $\Delta Vk-time$ has an accuracy rate of more than $97.1\% $for the early warning level of aircraft taxiing conflict. The accuracy rate of the warning level of confrontation conflicts is up to $98.1\%$, which is better than that of cross conflicts and rear-end conflicts. Experiments show that the proposed method has good accuracy in aircraft taxiing conflict warning level, and it can accurately and effectively predict different types and levels of conflict early warning after the controller issue command.

\subsubsection{Experimental Verification of Aircraft Taxiing Conflict Simulation Comparison}

Since there is currently no related research on aircraft taxiing conflict detection and early warning methods based on control command, the early warning methods of aircraft taxiing conflicts based on aircraft taxiing speed ranges established under different constraints are compared to verify the early warning effects of three types of taxiing conflicts. In the simulation comparison experiment, the aircraft taxiing conflict detection and early warning method in the $\Delta Vk-Pauta$ taxiing speed range in Section 3.1 is selected, which is recorded as the $\Delta Vk-Pauta$ taxiing conflict early warning comparison method. Based on the operation data of 50 groups of flights in the target airport on September 1, 2019, the $\Delta Vk-Pauta$ conflict early warning method was used to conduct simulation comparison experiments for the three conflicts. The Fig.19 shows the accuracy rates of the three conflict warning levels obtained in the experiment.

\begin{figure}[ht]
\centering\includegraphics[width=5in]{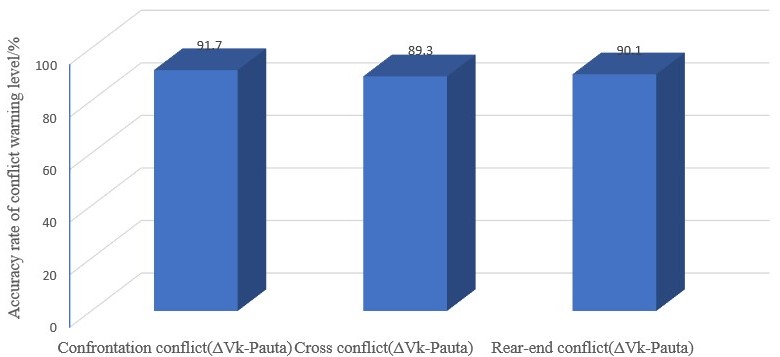}
\caption{Accuracy rates of three conflict warning levels in simulation comparison experiments}
\label{fig19}
\end{figure}

Through the analysis of the results in figure 19, it can be seen that the accuracy of the taxiing conflict early warning level of the aircraft taxiing conflict detection and early warning method based on the $\Delta Vk-Pauta$ taxiing speed interval is over $89.3\%$. Among them, the accuracy rate of the warning level of the cross conflict is the lowest at $89.3\%$. The accuracy rate of the warning level of confrontation conflicts is the highest, reaching $91.7\%$, which is better than that of cross conflict and rear-end conflict.

The accuracy rates of the three conflict warning levels obtained by the two methods of the $\Delta Vk-time$ taxiing conflict early warning method and the $\Delta Vk-Pauta$ conflict early warning method proposed in this paper are shown in table 14.

\clearpage

\begin{table}[h]\centering
\caption{Data representing three elements of aircraft type and taxiing speeds in the B-8 taxiway}%%%Table caption goes here
\label{table14}

\centering
    \begin{tabular}{m{3.5cm}<{\centering}m{3cm}<{\centering}m{3cm}<{\centering}m{3cm}<{\centering}}%%%The number of columns has to be defined here
    \hline
    Method &Accuracy of confrontation conflict warning level/\% &Accuracy of cross conflict warning level/\% &Accuracy of rear-end conflict warning level/\% \\
    \hline
$\Delta$Vk-time taxiing conflict early warning method &98.1 &97.1 &97.7 \\
 $\Delta$Vk-Pauta taxiing conflict early warning comparison method &91.7 &89.3 &90.1 \\
  \hline
\end{tabular}
\end{table}%%%End of the table

The experimental results in table 14 show that the effect of the aircraft taxiing conflict early warning method based on the $\Delta Vk-time$ taxiing speed interval is significantly better than that of the aircraft taxiing conflict early warning method based on the $\Delta Vk-Pauta$ taxiing speed interval. Among them, the accuracy rate of the early warning level of the confrontation conflict increased by $6.4\%$, the accuracy of the early warning level of the cross conflict increased by $7.8\%$, the accuracy of the early warning level of the rear-end conflict increased by $7.6\%$, and the accuracy of the three types of taxiing conflict early warning levels increased by $7.27\%$ on average. The results show that this paper not only proposes a new paradigm of aircraft taxiing conflict detection and early warning method, but also through simulation experiments and simulation comparison experiments show that the proposed method can accurately and effectively predict conflicts of different types and levels after the controller issues command. The early warning can better fit the changes of the real taxiing conflict alarm situation, and is closer to the real conflict situation during the taxiing process of the aircraft.

\subsection{Response Process of Aircraft Taxiing Conflict Early Warning Mechanism }

In recent years, according to the reports of airport accidents and incidents, incidents have shown an increasing trend, especially the frequency of serious incidents has increased. According to the theory of "Heinrich Safety Law", the continuous occurrence of serious incidents will inevitably lead to the occurrence of heavy casualties. To prevent the occurrence of major accidents, we must pay attention to the prewarning of accidents and incidents, and pay attention to the establishment of the response process of the accident early-warning mechanism.

According to modern human factors engineering theory, human error is the result, not the cause. Many accidents and incidents are rooted in design problems, that is, failure to implement "human-centered" design principles, or even inadequate system or mechanism design. The theory is applied in the field of aircraft taxiing conflict early warning. In the face of such a complex control environment, even the chief controller with rich control experience may make mistakes. Therefore, the research in this paper is not to discover human errors, but to explore how to avoid human errors or how to eliminate the hidden dangers and influences caused by human errors.

The purpose of this study is not to replace the controller. Both the method of aircraft taxiing process deduction proposed in Section 3 and the method of aircraft taxiing conflict detection and early warning proposed in Section 4 aim to correct the errors of control command in advance before the actual aircraft taxiing out/in and after the command given by controllers. Eliminate the impact of the controller's "wrong, forget, and leak", and only serve as an auxiliary decision-making support for the controller, so as to avoid accidents and incidents caused by human factors of the controller as much as possible.

Therefore, by following the predesigned aircraft taxiing conflict early warning and prevention countermeasures, and dynamically monitoring the aircraft taxiing conflict risk trend based on control command, this paper finally constructs an organically connected aircraft taxiing conflict early warning mechanism for risk identification, classification, early warning and intervention. The response process of the aircraft taxiing conflict early warning mechanism is proposed, as shown in figure 20. The controller can take different levels of response measures according to the different taxiing conflict risk warning levels in the response process of the aircraft taxiing conflict early warning mechanism, and can also transmit the results of the taxiing conflict risk early warning level to the pilot synchronously. The established response process of the aircraft taxiing conflict early warning mechanism can effectively improve the fault tolerance rate of the controller before the actual taxiing of the aircraft, and greatly enhance the stability and safety of the surface operation of the airport flight area.

\begin{figure}[ht]
\centering\includegraphics[width=5in]{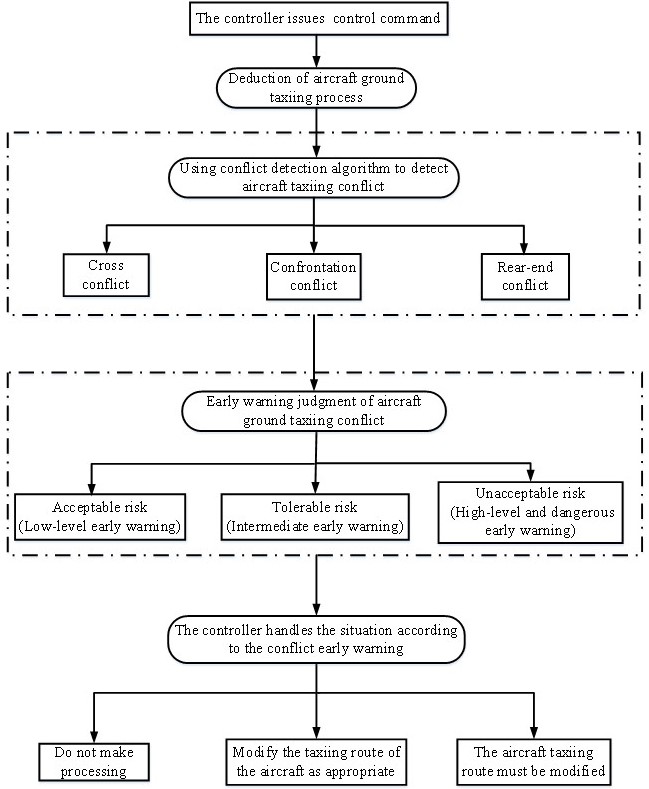}
\caption{Response flow chart of aircraft taxiing conflict early warning mechanism}
\label{fig20}
\end{figure}

For acceptable risk (low-level warning), the warning mechanism response process does not change the aircraft taxiing strategy issued by the controller, which is also consistent with the smooth aircraft taxiing movement target. For tolerable risk (intermediate early warning) and unacceptable risk (high-level and dangerous early warning), the taxiing route of the aircraft is modified as appropriate and the taxiing route of the aircraft must be modified. The requirements for the controller in the response process of the early warning mechanism are as follows. The re-issued control command must comply with China's "Administrative Measures for Civil Aviation Air Traffic Management Safety Assessment" Article 15: The risk acceptability criterion is an important basis for risk analysis. For the determination of the risk acceptability criterion, please refer to: (3) The overall risk is not expected to increase significantly after the change is implemented compared to before the implementation."

\section{Conclusion}

How to solve and deal with the problem of taxiing conflicts, how to improve the special situation handling ability and emergency early warning ability of the airport flight area, and how to eliminate the hidden dangers and influences caused by controller errors. In view of this situation, this paper proposes a method of aircraft taxiing deduction and conflict early warning based on control command information. The method deduces the taxiing process of the future aircraft based on the control command information, detects the aircraft taxiing conflict and obtains the conflict probability, and realizes early warning of different levels. And established the aircraft taxiing conflict early warning mechanism response process, according to different early warning levels to take different response measures.

Combined with the actual operation data of an airport in North China, the experiment shows that the correlation coefficient between the deduced taxiing process and the actual taxiing process is close to 1, indicating that the correlation is strong and the degree of correlation is good, which is a complete correlation. At the same time, the average deviation of the two is close to 0, indicating that the two have good consistency and high degree of fitting, and can better fit the change of the real value, which is closer to the real situation of the aircraft taxiing process. Experiments show that the accuracy of the three taxiing conflict early warning levels of the aircraft is over $98.1\%$. The proposed method can accurately and effectively predict different types and levels of conflict early warning after the controller issues command.

This research provides a solution to verify the correctness of the control command in advance after the controller issues the command, which can minimize the impact of "wrong, forget, and miss" by the controller. It can provide auxiliary decision sup-port for the controller, do a good job of supervision and reminding. To achieve the purpose of improving the control efficiency and the safety of the surface operation of the airport flight area. In addition, it can also provide better services for the upgrading of operating instruments and air traffic control equipment in the airport flight area.

The accurate deduction of aircraft ground taxiing process is related to the accurate identification of airport operation details. With the construction process of smart airport, scientific and technological means such as Internet of things and big data will be continuously applied to airport operation. The airport operation details such as off-block, pushback complete will be digitized. The airport operation process details will be accurately recorded and uploaded. Therefore, with the development of smart airport, the deduction and early warning of aircraft ground taxiing process will be more accurate, and should have broad application prospect. 

{\section*{Data Availability Statement}}

Some or all data, models, or code generated or used during the study are proprietary or confidential in nature and may only be provided with restrictions (ADS-B data).

{\section*{Acknowledgments}}

This work was supported by the Central University basic scientific research business fee special project of Civil Aviation University of China (3122019047) and the National Natural Science Foundation of China (52005500).

\end{sloppypar}
\end{document}